\documentclass[3p,times,sort&compress]{elsarticle}

\usepackage{graphicx}
\usepackage{epsfig,amsmath,amssymb,graphicx,pdflscape,color}
\usepackage{hyperref}
\usepackage{boondox-cal}
\usepackage[noabbrev]{cleveref}
\usepackage{siunitx}
\usepackage{color}
\usepackage[export]{adjustbox}

\crefname{section}{}{}
\crefname{equation}{}{}
\crefname{figure}{}{}
\crefname{table}{}{}
\crefname{appendix}{}{}
\crefname{chapter}{}{}

\def\cbl{\color{black}}
\def\cb{\color{black}}

\newcommand{\vect}[1]{\boldsymbol{#1}}
\DeclareRobustCommand{\e}{\relax\ifmmode\mathrm{e}\else\error\fi}
\usepackage[title]{appendix}
\begin{document}

\begin{frontmatter}
\title{Survival, extinction, and interface stability in a two--phase moving boundary model of biological invasion}

\author[qut]{Matthew J. Simpson \corref{cor1}}
\author[qut]{Nizhum Rahman}
\author[qut]{Scott W. McCue}
\author[UniSA]{Alexander K.Y. Tam}
\address[qut]{School of Mathematical Sciences, Queensland University of Technology, Brisbane, Australia.}
\address[UniSA]{UniSA STEM, Mawson Lakes Campus, The University of South Australia, Mawson Lakes, SA 5095, Australia.}
\cortext[cor1]{Corresponding author: matthew.simpson@qut.edu.au}

\begin{abstract}
We consider a moving boundary mathematical model of biological invasion. The model describes the spatiotemporal evolution of two \cbl adjacent \cb populations:  each population undergoes linear diffusion and logistic growth, and the boundary between the two populations evolves according to a two--phase Stefan condition. This mathematical model describes situations where one population invades into regions occupied by the other population, such as the spreading of a malignant tumour into surrounding tissues. Full time--dependent numerical solutions are obtained using a level--set numerical method. We use these numerical solutions to explore several properties of the model including: (i) survival and extinction of one population initially surrounded by the other; and (ii) linear stability of the moving front boundary in the context of a travelling wave solution subjected to transverse perturbations.  Overall, we show that many features of the well--studied one--phase single population analogue of this model can be very different in the more realistic two--phase setting.  These results are important because realistic examples of biological invasion involve interactions between multiple populations and so great care should be taken when extrapolating predictions from a one--phase single population model to cases for which multiple populations are present.  Open source Julia--based software is available on \href{https://github.com/alex-tam/TwoPhaseInvasion}{GitHub} to replicate all results in this study.
\end{abstract}

\begin{keyword}
Reaction--diffusion; Travelling wave; Linear stability; Level--set method; Stefan problem.
\end{keyword}
\end{frontmatter}

\newpage
\section{Introduction}\label{sec:Intro}

\cbl
Mathematical models are routinely used to study biological population phenomena~\cite{Murray2002,Kot2003,Edelstein2005}, with a key focus being the long term survival or extinction of a population.   One of the most commonly used mathematical models of biological population growth is the logistic growth model that describes the growth of a population with density $C(t)$~\cite{Murray2002,Kot2003,Edelstein2005}.    For small densities the logistic growth model predicts approximately exponential growth, and as the density increases the net growth rate slows as the density reaches carrying capacity density, $C(t) \to K$ as $t \to \infty$.  One limitation of the logistic growth model is that it predicts that all populations survive and grow to eventually reach  the carrying capacity density  $K$.  This means that standard applications of the logistic growth model cannot be used to study biological extinction, which is a clear limitation of this model.  A simple way of overcoming this limitation is to extend the quadratic logistic growth model to a cubic model that exhibits  bistability, which is often called an Allee effect~\cite{Keitt2001,Taylor2005,Barton2011}.  Mathematical models including an Allee effect typically involve an unstable equilibrium density $0 < A < K$, such that when $C(0) > A$ the population density eventually approaches $K$, whereas when $C(0) < A$ the population will eventually go extinct~\cite{Fadai2020,Surendran2020}.   When reaction--diffusion equations are implemented with Allee--type source terms, other features impact long--time survival or extinction.  For example, both the spatial extent~\cite{Lewis1993} and the shape of the initial distributions~\cite{Li2022} can determine whether the population survives and spreads, or goes extinct for these models.

While the Allee effect, and generalisations thereof are often used in mathematical models describing the dynamics of homogeneous biological populations, another approach to study population survival or extinction is to work within a framework of modelling two or more interacting populations~\cite{Murray2002,Kot2003,Edelstein2005}, where each population can have different properties and can compete with each other for space and/or nutrients.  These kinds of mathematical models, often based on the well--known Lotka--Volterra model~\cite{Murray2002,Kot2003,Edelstein2005}, can predict long--term survival of one population and extinction of the other, or long--term co--existence of both populations~\cite{Cantrell1998,Maciel2013,Fussell2019}  An important feature of standard applications of Lotka--Volterra--type models is an implicit assumption that the system is sufficiently well--mixed so that both populations can occupy the same region of space so that individuals within each population and mix and directly compete with each other, for example through predator--prey type interactions.

\begin{figure}
    \centering
    \includegraphics[width=1.0\textwidth]{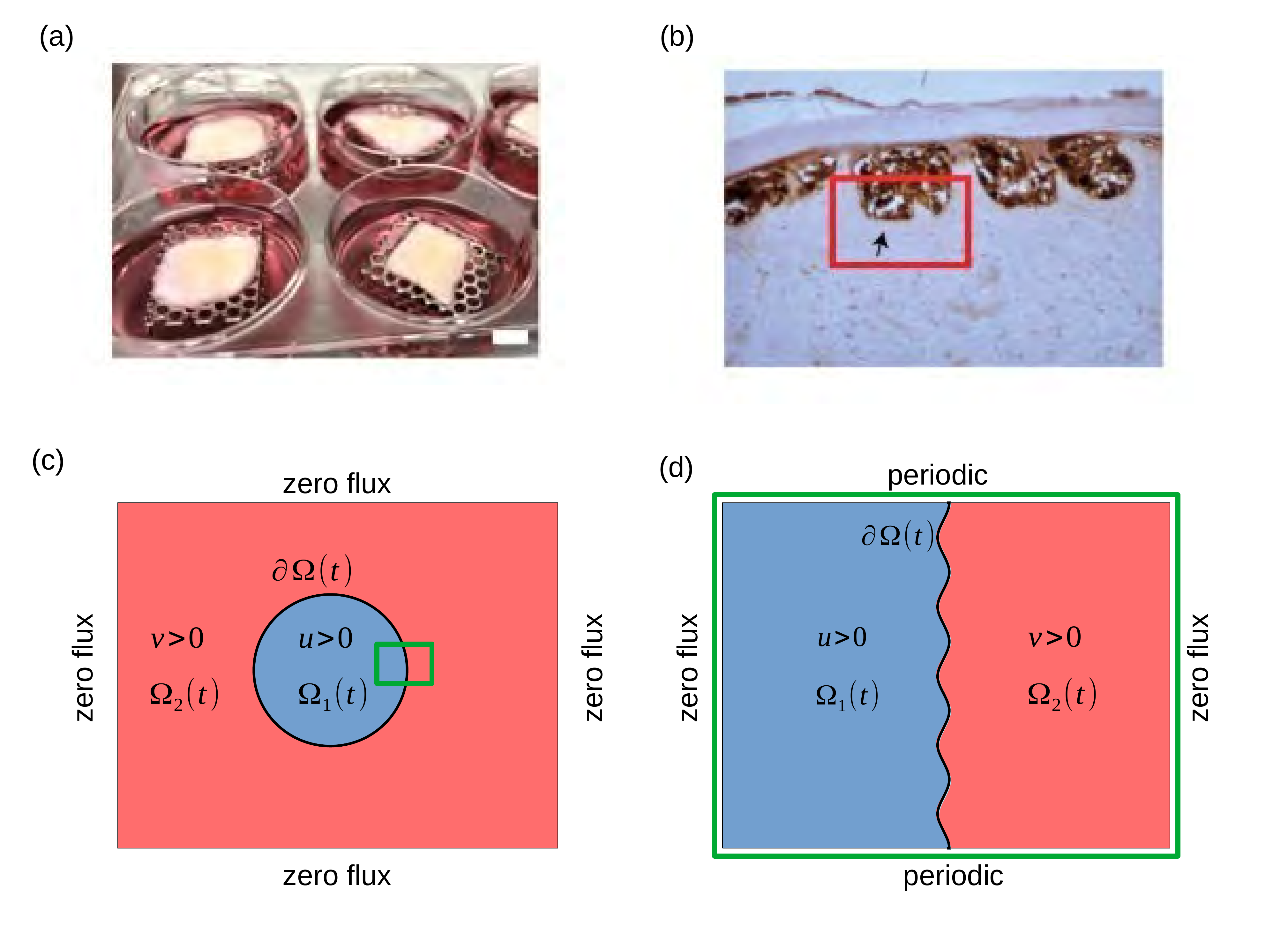}
    \caption{Experimental motivation and mathematical modelling schematics illustrating a two--phase invasion model with population densities $u$ and $v$. (a) Melanoma invasion experiments showing donated human skin tissues maintained at an air--liquid interface to study the invasion of melanoma cells into the surrounding skin tissues.  (b) Cross section through the skin tissues showing the downward invasion of melanoma cells (brown) into the surrounding skin tissues (pink).  The interface between the invading melanoma population and the receding skin tissue is highlighted in the red rectangle. All images are reproduced from~\cite{Haridas2017} with permission.  (c) Mathematical model schematic used to study  the survival or extinction of population $u$ that is initially surrounded by population $v$.  (d) Mathematical model schematic to study the stability of the interface separating population $u$ from population $v$ in the context of a travelling wave solution subjected to transverse perturbations.  As in (c), population $u$ occupies $\Omega_1(t)$, population $v$ occupies $\Omega_2(t)$ and the moving boundary separating the two populations where $u=v=0$ is denoted $\partial \Omega(t)$ (solid black curve).  In (c)--(d) the computational domain $\mathcal{D}$ is rectangular and appropriate boundary conditions along the boundaries of $\mathcal{D}$ are labelled.}
    \label{Figure_1}
\end{figure}

 The focus of the present work is very different to either an Allee--type or Lotka--Volterra--type model.  Here we are interested in two adjacent, unmixed biological populations separated by a sharp interface.  This modelling framework is motivated by experimental images of tumour growth, where a population of motile and proliferative cancer cells invades into a surrounding healthy tissue that is composed of a different population of motile and proliferative cells.  Experimental images reported by Gatenby and Gawlinski~\cite{Gatenby1996} show an invading population of tumour cells that is surrounded by a population of normal non-tumour tissue, with a visually distinct cell--free gap between the two populations.  The experimental image in Figure \ref{Figure_1}(a)--(b) shows a similar situation involving a population of malignant melanoma cells invading into a population of surrounding skin cells.  Here, it is relevant to use mathematical models of biological invasion to explore conditions that govern the long--term survival or extinction of the melanoma cells.  Since Gatenby and Gawlinski first published their clinical observations of the gap between the invasive and receding populations, there have been several different types of mathematical models proposed to study this kind of phenomenon~\cite{Schofield2011,Browning2019}.  Importantly, this kind of situation is different to a standard application of a Lotka--Volterra type model since here the two populations are unmixed, and do not occupy the same spatial location. \cb

Reaction--diffusion equations are routinely used in mathematical biology and mathematical ecology to study population invasion~\cite{Murray2002,Kot2003,Edelstein2005}.  Typically, these models involve a diffusion term to represent migration of individuals within the population, and a logistic source term to represent carrying capacity--limited proliferation.  The Fisher--Kolmogorov model~\cite{Fisher1937,Kolmogorov1937}, and generalisations thereof, are prototype reaction--diffusion models that are extremely well--studied and broadly applied in ecology~\cite{Skellam1951} and cell biology~\cite{Maini2004,Sengers2007,Jin2016,Jin2018}.  A key limitation of reaction--diffusion models that incorporate linear diffusion, however, is that the solutions of these models lack a well--defined front, which is often present in practical applications~\cite{Maini2004,McCue2019,ElHachem2019}.  One way to address this shortcoming is to reformulate these classical reaction--diffusion models as moving boundary problems with a Stefan--like condition at the moving boundary where the population density vanishes~\cite{Crank1987,Gupta2017}.  Combining Fisher--Kolmogorov style models with a moving boundary condition gives rise to a family of very interesting mathematical models that have been called \textit{Fisher--Stefan} models~\cite{ElHachem2019,Du2010}.  While moving boundary conditions are routinely employed in mathematical models of heat and mass transfer, as well as mathematical models of fluid mechanics processes~\cite{BrosaPlanella2019,BrosaPlanella2021,Mitchell2009,Mitchell2010,Dalwadi2020}, these approaches are less widely adopted to study population biology processes~\cite{Ward1997,Gaffney1999,Shuttleworth2019,ElHachem2021}.  This is despite some clear advantages, such as allowing us to study population invasion with well--defined fronts, as well as examining biological invasion and recession within a relatively simple modelling framework.

Fisher--Stefan type partial differential equation (PDE) models have been studied extensively in one--dimensional Cartesian coordinates~\cite{Du2010,ElHachem2019} or in radially symmetric geometries with one spatial independent variable~\cite{Simpson2020}.  Working in one dimension is useful because it allows us to study both the \textit{spreading--vanishing dichotomy}~\cite{Du2010,ElHachem2019}, as well as long--time travelling wave solutions associated with these models~\cite{Du2010,ElHachem2019}.   Most previous studies about travelling wave solutions of Fisher--Stefan models focus on one--phase single--species problems that describe the evolution of a single population with density $u$~\cite{Du2010,ElHachem2019}.  Simple models like this can be very instructive because it is intuitive to think about a population with density $u$ that invades into an adjacent empty region.  If, however, we are interested in a more realistic scenarios where invasion often occurs by one population displacing another \cbl adjacent population, \cb a single species model is insufficient.  To address this limitation El-Hachem et al.~\cite{ElHachem2020} proposed an extended two--phase Fisher--Stefan model that describes the evolution of two \cbl adjacent, unmixed populations, \cb with densities $u$ and $v$, where both populations undergo linear diffusion and logistic growth and the interface between the two populations, where $u=v=0$, moves according to a two--phase Stefan condition.  \cbl Setting $u=v=0$ is a simple way to model the presence of the experimentally--observed cell--free gap. \cb This two--phase extension can be used to study the invasion of one population into another surrounding population or adjacent population, such as describing motile and proliferative melanoma cancer cells that invade into surrounding motile and proliferative skin cells.  This scenario, where the two biological populations are separated by a sharp front, is illustrated in Fig.~\ref{Figure_1}(a)--(b) where experimental images show a population of melanoma cells invading into a population of surrounding skin cells~\cite{Haridas2017}.

\newpage
Previously, El-Hachem et al.~\cite{ElHachem2020} showed that their one--dimensional two--phase moving boundary model of biological invasion leads to a family of long--time travelling wave solutions.  El-Hachem et al.~\cite{ElHachem2020} studied the properties of these travelling wave solutions using a combination of full time--dependent numerical solutions, phase plane analysis and perturbation methods.  This model is biologically interesting because it leads to sharp--fronted travelling wave solutions, moving with speed $c$, that describes a range of scenarios including situations where: \cbl (i) population $u$ invades into population $v$; (ii) population $v$ invades into population $u$; and (iii) a steady state profile with $c=0$ where neither $u$ or $v$ invades into the other population. \cb

While Fisher--Stefan--type models have been relatively well--studied in one spatial dimension, much less attention has been paid to these models in more general geometries.  In this work we re--visit the two--phase Fisher--Stefan type model introduced by El-Hachem et al.~\cite{ElHachem2020} in two dimensions.  This setting is appropriate for studying more general two--species biological invasion phenomena, including the experimental scenarios shown in Figure~\ref{Figure_1}(a)--(b) illustrating the sharp--fronted invasion of a population of motile and proliferative melanoma cells within a population of surrounding skin cells. We use a level--set numerical method to obtain time--dependent numerical solutions of the PDE model and we revisit the spreading--vanishing dichotomy in the two--phase setting where one population is initially surrounded by another population, such as the schematic in Figure~\ref{Figure_1}(c). In summary, we show that the distinction between long--term survival or long--term extinction is more subtle in the two--phase model rather than the well--established results in the one--phase case~\cite{Du2010,Simpson2020,Tam2022}.  We also re--visit travelling wave solutions in the two--phase model and explore the stability of travelling wave solutions subject to a small transverse perturbation~\cite{Tam2023}, as illustrated schematically in Figure~\ref{Figure_1}(d).  We show that travelling wave solutions can be linearly unstable or linearly stable depending on the parameters in the mathematical model, and demonstrate that our short--time linear stability analysis is consistent with full time--dependent numerical solutions of the governing system of PDEs.  Overall we show that the distinction between linearly stable and linearly unstable travelling wave solutions is subtly different to previous results where we explored the stability of travelling wave solutions in one  phase only.

\newpage

\section{Mathematical model}\label{sec:mathmodel}
\subsection{Governing Equations}\label{sec:governingequations}
We consider the two--phase moving boundary model describing the evolution of two populations with dimensional density  $\bar{u}(\bar{x},\bar{y},\bar{t})\;[\text{cells}/L^2]$ and $\bar{v}(\bar{x},\bar{y},\bar{t})\;[\text{cells}/L^2]$,
\begin{subequations}
\label{eq:Dimensional_model}
\begin{align}
        \frac{\partial \bar{u}}{\partial \bar{t}}&=\bar{D}_u\bar{\nabla}^2 \bar{u}+\; \bar{\lambda}_u \bar{u}\left(1-\frac{\bar{u}}{\bar{K}_u}\right) \;\; \text{on}   \;\;    \cbl \bar{\Omega}_1 (\bar{t}), \cb \label{Sub_eq:dimensional_population_u_diffusion} \\
       \frac{\partial \bar{v}}{\partial \bar{t}}&=\bar{D}_v\;\bar{\nabla}^2 \bar{v}+\;\bar{\lambda}_v \bar{v}\left(1-\frac{\bar{v}}{\bar{K}_v}\right) \;\; \text{on}   \;\;   \cbl \bar{\Omega}_2 (\bar{t}), \cb \label{Sub_eq:dimensional_population_v_diffusion}\\
 	\bar{u}(\bar{x},\bar{y},\bar{t})&=\bar{v}(\bar{x},\bar{y},\bar{t})=0 \;\; \text{on}   \;\; \cbl \partial \bar{\Omega} (\bar{t}), \cb \label{Sub_eq:dimensional_population_density_at_interface}\\
	\bar{\mathcal{v}}_\textrm{n}&=-\bar{\kappa}_u \bar{\nabla} \bar{u} \cdot \hat{n}-\bar{\kappa_v}\bar{\nabla} \bar{v} \cdot \hat{n}\;\;  \text{on}   \;\;  \cbl \partial \bar{\Omega} (\bar{t}), \cb \label{Sub_eq:dimensional_velocity_at_interface}\\
	\bar{u}(\bar{x},\bar{y},0)&={\bar{U}}(\bar{x},\bar{y}) \;\; \text{on}   \;\; \cbl \bar{\Omega}_1 (0), \cb \label{Sub_eq:dimensional_initial_popution_u}\\
	\bar{v}(\bar{x},\bar{y},0)&={\bar{V}}(\bar{x},\bar{y}) \;\; \text{on}   \;\; \cbl \bar{\Omega}_2 (0). \cb \label{Sub_eq:dimensional_initial_popution_v}
\end{align}
\end{subequations}
In this model, population $\bar{u}$ undergoes linear diffusion with diffusivity $\bar{D}_u$ $[L^2/T]$, logistic proliferation with proliferation rate $\bar{\lambda}_u$ $[1/T]$, has carrying capacity density $\bar{K}_u$ $[\text{cells}/L^2]> 0$, and occupies $\Omega_1 (\bar{t})$.  Similarly, population $\bar{v}$ undergoes linear diffusion with diffusivity $\bar{D}_v$ $[L^2/T]$, logistic proliferation with proliferation rate $\bar{\lambda}_v$ $[1/T]$, has carrying capacity density $\bar{K}_v$ $[\text{cells}/L^2]> 0$, and occupies  $\Omega_2 (\bar{t})$.  The density of both populations vanishes on the boundary between the two populations, $\partial \Omega(\bar{t})$, and the normal velocity of this interface $\bar{\mathcal{v}}_\textrm{n}$ $[L/T]$ is given by a two--phase Stefan condition that involves two constants, $\bar{\kappa}_u$ $\left[L^4/(T \times \text{cells})\right]$ and $\bar{\kappa}_v$ $\left[L^4/(T \times \text{cells})\right]$~\cite{ElHachem2020}, and $\hat{n}$ is the unit normal along the moving boundary.  \cbl The moving boundary condition (\ref{Sub_eq:dimensional_population_density_at_interface}) requires that both densities vanish along the interface, which is consistent with experimental observations of a small cell--free gap between the invading and receding populations~\cite{Gatenby1996}. \cb

Throughout this work all dimensional quantities are written with an overbar, and analogous dimensionless quantities are written as regular variables without the overbar. To simplify the mathematical model we introduce non--dimensional dependent variables $u = \bar{u} / \bar{K}_u$ and $v = \bar{v} / \bar{K}_v$, and non--dimensional independent variables $x = \bar{x}\sqrt{\bar{\lambda}_u/\bar{D}_u}$, $y = \bar{y}\sqrt{\bar{\lambda}_u/\bar{D}_u}$ and $t = \bar{t}\bar{\lambda}_u$ to give
\begin{subequations}
\label{eq:Non-dimensional_model}
\begin{align}
        \frac{\partial u}{\partial t}&=\nabla^2u+\; u(1-u) \;\; \text{on}   \;\;   \Omega_1 (t), \label{Sub_eq:non-dimensional_population_u_diffusion} \\
       \frac{\partial v}{\partial t}&=D\;\nabla^2v+\;\lambda v(1-v), \;\; \text{on}   \;\;   \Omega_2 (t),\label{Sub_eq:non-dimensional_population_v_diffusion}\\
 	u(x,y,t)&=v(x,y,t)=0 \;\; \text{on}   \;\;  \partial \Omega (t),\label{Sub_eq:non-dimensional_population_density_at_interface}\\
	\mathcal{v}_\textrm{n}&=-\kappa_u\nabla u \cdot \hat{n}-\kappa_v\nabla v \cdot \hat{n}\;\;  \text{on}   \;\;  \partial \Omega (t),\label{Sub_eq:non-dimensional_velocity_at_interface}\\
	u(x,y,0)&={U}(x,y) \;\; \text{on}   \;\; \Omega_1 (0),\label{Sub_eq:non-dimensional_initial_popution_u}\\
	v(x,y,0)&={V}(x,y) \;\; \text{on}   \;\; \Omega_2 (0).\label{Sub_eq:non-dimensional_initial_popution_v}\;
\end{align}
\end{subequations}
The nondimensional model (\ref{eq:Non-dimensional_model}) has four parameters, namely
\begin{equation}
    \label{eq:non-dimensional_parameters}
    D=\frac{\bar{D}_v}{\bar{D}_u},\quad \lambda=\frac{\bar{\lambda}_v}{\bar{\lambda}_u}, \quad \kappa_u=\frac{ \bar{\kappa}_u\bar{K}_u}{\bar{D}_u}, \quad \kappa_v=\frac{\bar{\kappa}_v\bar{K}_v}{\bar{D}_u},
\end{equation}
where $D$ is a relative diffusivity, $\lambda$ is a relative proliferation rate, and $\kappa_u$ and $\kappa_v$ are non--dimensional constants that are proportional to the dimensional coefficients in the dimensional moving boundary condition  (\ref{Sub_eq:dimensional_velocity_at_interface}).

\begin{figure}
\centering
\includegraphics[width=1.0\textwidth]{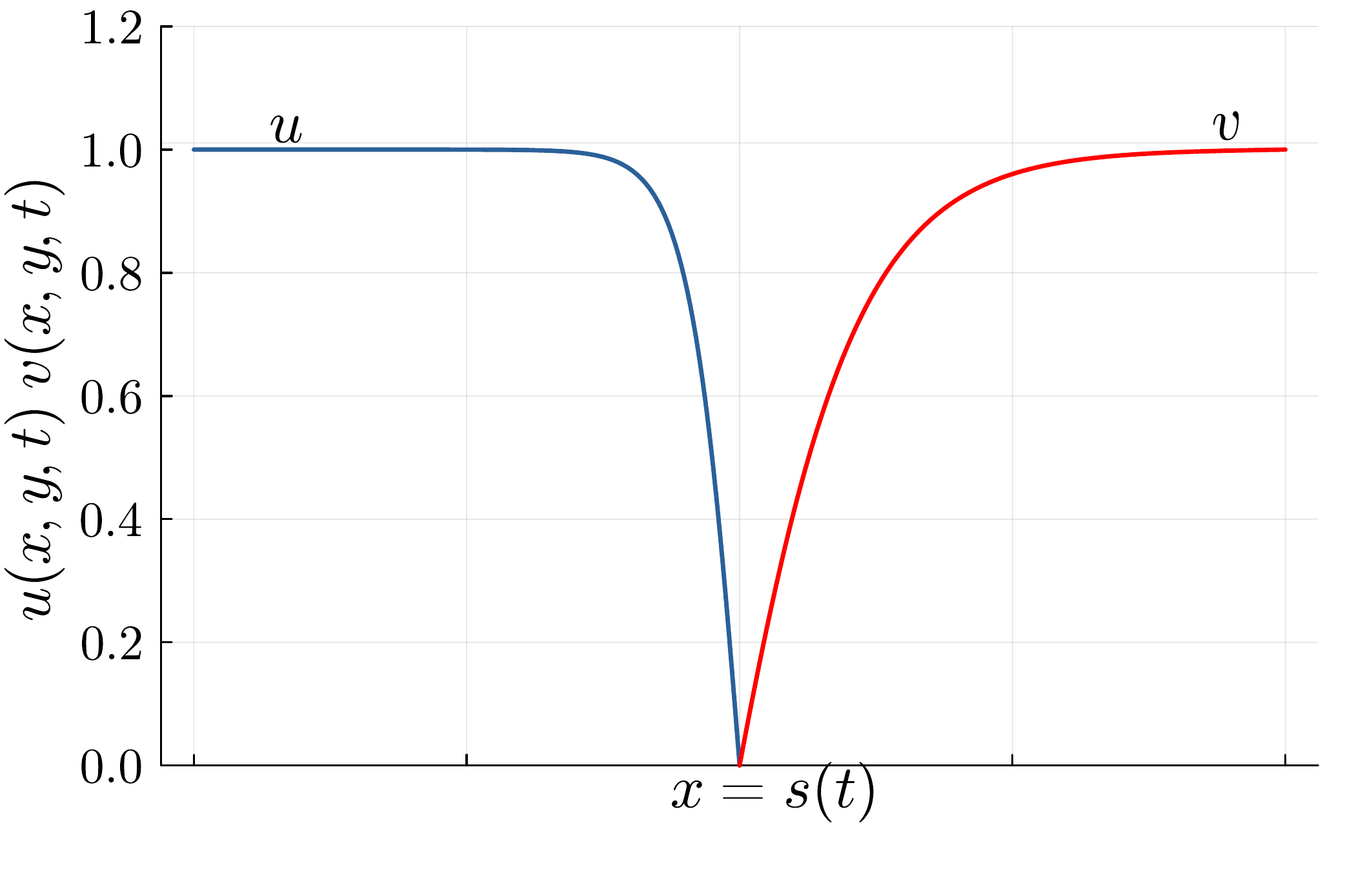}
\caption{Interface schematic showing $u(x,y,t)$ (solid blue), $v(x,y,t)$ (solid red), and the location of the interface at $x = s(t)$.  Here profiles are shown along some horizontal line $y=Y$.}
\label{Figure_2}
\end{figure}

\cbl As we will show in Section \ref{sec: survival}, solutions of this non-dimensional model give rise to moving interfaces as depicted schematically in Figure \ref{Figure_2} where we show a one--dimensional interface at $x=s(t)$ for clarity.  At the interface, $x = s(t)$, we have local loss of $u$ and $v$ since the diffusive flux of $u$ is in the positive $x$--direction, and the diffusive flux of $v$ is in the negative $x$--direction owing to Fick's first law.  The interface velocity is $\textrm{d}s(t) / \textrm{d} t = -\kappa_u \partial u / \partial x - \kappa_v \partial v / \partial x$ at $x = s(t)$, so that the choices of $\kappa_u$ and $\kappa_v$ together with the shapes of the density profiles at $x=s(t)$ determine the direction that $s(t)$ moves.   If the choices of $\kappa_u$ and $\kappa_v$ leads to $\textrm{d}s(t) / \textrm{d} t >0$, then at some location $X$ close to $s(t)$ we will eventually have $u(X,t) = 1$ and $v(X,t) =0$, which we can interpret as a local conversion of $v$ into $u$, as population $u$ invades and population $v$ recedes.  Conversely, if the choices of $\kappa_u$ and $\kappa_v$ leads to $\textrm{d}s(t) / \textrm{d} t <0$, then at some location $X$ close to $s(t)$ we will eventually have $u(X,t) = 0$ and $v(X,t) =1$, which we can interpret as a local conversion of $u$ into $v$, as population $v$ invades and population $u$ recedes.  This interpretation is consistent with clinical observations of cellular invasion showing that the invading population evolves to occupy new regions of space and grows at the expense of the receding population~\cite{Gatenby1996}.  \cb

\subsection{Numerical Method}\label{sec:numericalmethods}
We study the solution of the system~\eqref{eq:Non-dimensional_model} on a two--dimensional rectangular domain $\mathcal{D}$ by implementing a level--set numerical method~\cite{Tam2023,Sethian1999,Osher2003}.  Briefly, the moving boundary is embedded within $\mathcal{D}$ as the zero level--set of the signed distance function $\phi(x,y,t)$ which satisfies the Eikonal equation $\mid \nabla(\phi) \mid = 1$.  The position of the interface is implicitly defined by
\begin{equation}\label{eq:Interface}
    \partial \Omega(t)=\{(x,y) \mid \phi(x,y,t)=0 \},
\end{equation}
where $\phi(x,y,t)$ is defined throughout $\mathcal{D}$ with the property that $\phi<0$ on $\Omega_1(t)$ where $u(x,y,t)$ is present, and $\phi>0$ on $\Omega_2(t)$ where $v(x,y,t)$ is present.  To ensure that $\phi=0$ is maintained on the interface $\partial \Omega(t)$,  $\phi$ evolves according to the level--set equation
\begin{equation}\label{eq:Velocity_field}
    \frac{\partial \phi}{\partial t}= -F\mid\nabla\phi\mid,
\end{equation}
where $F(x,y,t)$ is an extension velocity field that is defined on $\mathcal{D}$ with the property that $F = \mathcal{v}_\textrm{n}$ on $\partial \Omega(t)$.  Following Tam and Simpson~\cite{Tam2023} we reinitialize $\phi$ after every time step~\cite{Osher2003} to ensure that the signed--distance property is maintained.

Within this framework, we re--write (\ref{eq:Non-dimensional_model}) as
\begin{subequations}
\label{eq:Level_set_model}
\begin{align}
        \frac{\partial u}{\partial t}&=\nabla^2u+\; u(1-u) \;\; \text{where}   \;\;   \phi(x,y,t)<0, \label{Sub_eq:level_set_population_u_diffusion} \\
       \frac{\partial v}{\partial t}&=D\;\nabla^2v+\;\lambda v(1-v) \;\; \text{where}   \;\;   \phi(x,y,t)>0,\label{Sub_eq:level_set_population_v_diffusion}\\
       \frac{\partial \phi}{\partial t}&= -F\mid\nabla\phi\mid \;\;\text{on}   \;\;   \mathcal{D},\label{Sub_eq:level_set_velocity_field}\\
 	u(x,y,t)&=v(x,y,t)=0 \;\; \text{where}   \;\;  \phi(x,y,t)=0,\label{Sub_eq:level_set_population_density_at_interface}\\
	F&=-\kappa_u\nabla u \cdot \frac{\nabla \phi}{|\nabla \phi|}-\kappa_v\nabla v \cdot \frac{\nabla \phi}{|\nabla \phi|}\;\; \text{where}   \;\;  \phi(x,y,t)=0, \label{Sub_eq:level_set_velocity_at_interface}\\
	u(x,y,0)&={U}(x,y) \;\; \text{on}   \;\;  \phi(x,y,0)<0,\label{Sub_eq:level_set_initial_popution_u}\\
	v(x,y,0)&={V}(x,y) \;\; \text{on}   \;\;  \phi(x,y,0)>0.\label{Sub_eq:level_set_initial_popution_v}\;
\end{align}
\end{subequations}
We discretize $\mathcal{D}$ using a standard uniformly--spaced square finite--difference mesh, and we discretize the terms on the right--hand side of Equations (\ref{Sub_eq:level_set_population_u_diffusion})--(\ref{Sub_eq:level_set_population_u_diffusion}) using standard second--order finite--difference stencils, taking care to deal with cases for which the moving boundary lies between mesh points using interpolation~\cite{Tam2022}.  To solve Equation (\ref{Sub_eq:level_set_velocity_field}) we estimate $F$ by computing the normal velocity $\mathcal{v}_\textrm{n}$ using a second--order finite--difference approximation and orthogonal extrapolation~\cite{Osher2003,Aslam2004}; the resulting hyperbolic PDE is then discretized using a high resolution central scheme~\cite{Simpson2005} and the resulting system of coupled ordinary differential equations (ODE) are solved numerically using standard ODE solvers in Julia~\cite{Rackauckas2017}.   All numerical results in this work make use of a square mesh to discretize $\mathcal{D}$ with mesh spacing $h=0.1$.  Automatic time stepping within Julia's ODE solvers are employed and we save the results at constant time intervals of duration $\Delta t = 0.01$ and use the saved data to re--solve the level-set equations and re-initialise $\phi$.  As we will explain in Section \ref{sec: stability}, we first tested the accuracy of the level--set numerical method by solving a number of problems that lead to tractable travelling wave solutions as analysed by  El-Hachem et al.~\cite{ElHachem2020}, as well as checking our full two--dimensional numerical solutions match various radially symmetric problems reported previously by Simpson~\cite{Simpson2020}.  These initial test cases (not shown) confirm the veracity of our level--set method and for our choice of $h$ and $\Delta t$.  \cbl Furthermore, for all stable problems considered in this work  our time--dependent PDE solutions are grid-independent, and our open source code can be used to test for grid-independence when solving new problems with different choices of initial conditions and parameter values. \cb

Before we present and discuss numerical results it is worth noting that we will use our numerical solution of the system (\ref{eq:Level_set_model}) to study two different types of problems:  (i) Specifying values of $D$, $\lambda$, $\kappa_u$ and $\kappa_v$, as well as initial conditions $U(x,y)$, $V(x,y)$ and $\phi(x,y,0)$, we obtain full time--dependent PDE solutions for the two--phase invasion model; and, (ii) Setting $V(x,y)=0$ allows us to study a one--phase invasion problem where we only need to specify $U(x,y)$, $\kappa_u$ and $\phi(x,y,0)$.  As we will demonstrate in Section \ref{sec: survival}, it is convenient to use the same level--set numerical algorithm to generate and compare solutions of both one-- and two--phase invasion models for the same parameter values and initial condition.

\section{Results and Discussion I: Survival and Extinction} \label{sec: survival}

\subsection{Preamble: Survival and extinction for one--phase Fisher--Stefan problems} \label{sec: survival preamble}
Before we consider population survival and extinction in the two--phase invasion model, it is instructive to recall some established results for the analogous  one--phase invasion model, including the distinction between long--term survival and extinction, often called the spreading--vanishing dichotomy~\cite{Du2010,ElHachem2019}.  We begin by briefly summarising known results for the non--dimensional one--phase Fisher--Stefan model in a radially--symmetric coordinate system
\begin{subequations}  \label{eq:spreadingvanishing}
\begin{align}
\dfrac{\partial u}{\partial t} &= \dfrac{1}{r^{d-1}} \dfrac{\partial}{\partial r}\left(r^{d-1} \dfrac{\partial u}{\partial r} \right) + u(1-u), \quad 0 < r < R(t), \\
\dfrac{\partial u}{\partial r} &=0, \quad \textrm{on} \quad  r=0, \\
u &=0, \quad \dfrac{\textrm{d}R(t)}{\textrm{d}t} = -\kappa \dfrac{\partial u}{\partial r}, \quad \textrm{on} \quad r = R(t),
\end{align}
\end{subequations}
where we have written the model in terms of one spatial variable $r$ that is relevant for studying solutions of the model on a line $(d=1)$, a disc $(d=2)$ or a sphere $(d=3)$.  Note that in the non--dimensional one--phase model there is only one parameter in the moving boundary condition, $\kappa$, whereas in the two--phase model there are two parameters, $\kappa_u$ and $\kappa_v$.  The spreading--vanishing dichotomy refers to the fact that the solution of the moving boundary problem (\ref{eq:spreadingvanishing}) for certain initial conditions and values of $\kappa$ leads to eventual spreading of $u$ in the form of a travelling wave as $t \to \infty$, whereas other initial conditions and values of $\kappa$ lead to extinction, namely $u \to 0^+$ as $t \to \infty$~\cite{Du2010}.  A key feature of the spreading--vanishing dichotomy is that there exists a critical radius $R_{\textrm{c}}$, and solutions of Equation (\ref{eq:spreadingvanishing}) that evolve such that $R(t) > R_{\textrm{c}}$ always lead to long--time spreading, whereas if $R(t)$ never exceeds $R_{\textrm{c}}$ the population becomes extinct as $t \to \infty$.  Precise values of $R_{\textrm{c}}$ are well known; for example, $R_{\textrm{c}} = \pi / 2$ in one--dimensional Cartesian coordinates $(d=1)$, $R_{\textrm{c}}$ on a disc $(d=2)$ is the first zero of the zeroth--order Bessel function of the first kind giving $R_{\textrm{c}} \approx 2.4048$, and $R_{\textrm{c}} = \pi$ on a sphere $(d=3)$~\cite{Simpson2020}.  Tam and Simpson~\cite{Tam2022} recently derived analogous results for two--dimensional problems without radial symmetry, showing that square--shaped populations survive if the side length exceeds $\pi \sqrt{2}$, and rectangular--shaped populations survive if the area of the rectangle exceeds $\pi \sqrt{W^2 + H^2}$, where $W$ and $H$ is the width and height of the rectangle, respectively.   Intuitive derivations of these critical lengths have been reported previously, and can be derived using linearisation and separation of variables~\cite{Simpson2020,Tam2022}.

In summary, for the one--phase Fisher--Stefan invasion model, long--term survival is associated with populations that occupy a sufficiently large region, whereas extinction is associated with populations that do not occupy a sufficiently large region.  These results makes intuitive sense because extinction arises when the outward flux of the population at the moving boundary exceeds the total population gain through the source term.  Since the source term acts to increase the population density at all locations where $0 < u < 1$, the total population increases whenever the population occupies a sufficiently large region, whereas if the spatial extent of the population is not sufficiently large the outward flux at the moving boundary exceeds the population gain through the source term.  Indeed, deriving expressions for the critical radius~\cite{Simpson2020} and critical lengths~\cite{Tam2022} involves equating the rate of population loss at the moving boundary with the rate of population gain through the action of the source term; the resulting values of  $R_{\textrm{c}}$ have been numerically verified~\cite{ElHachem2019,Simpson2020,Tam2022}.

\subsection{Survival and extinction for the two--phase invasion model}
\begin{figure}
    \centering
    \includegraphics[width=1.0\textwidth]{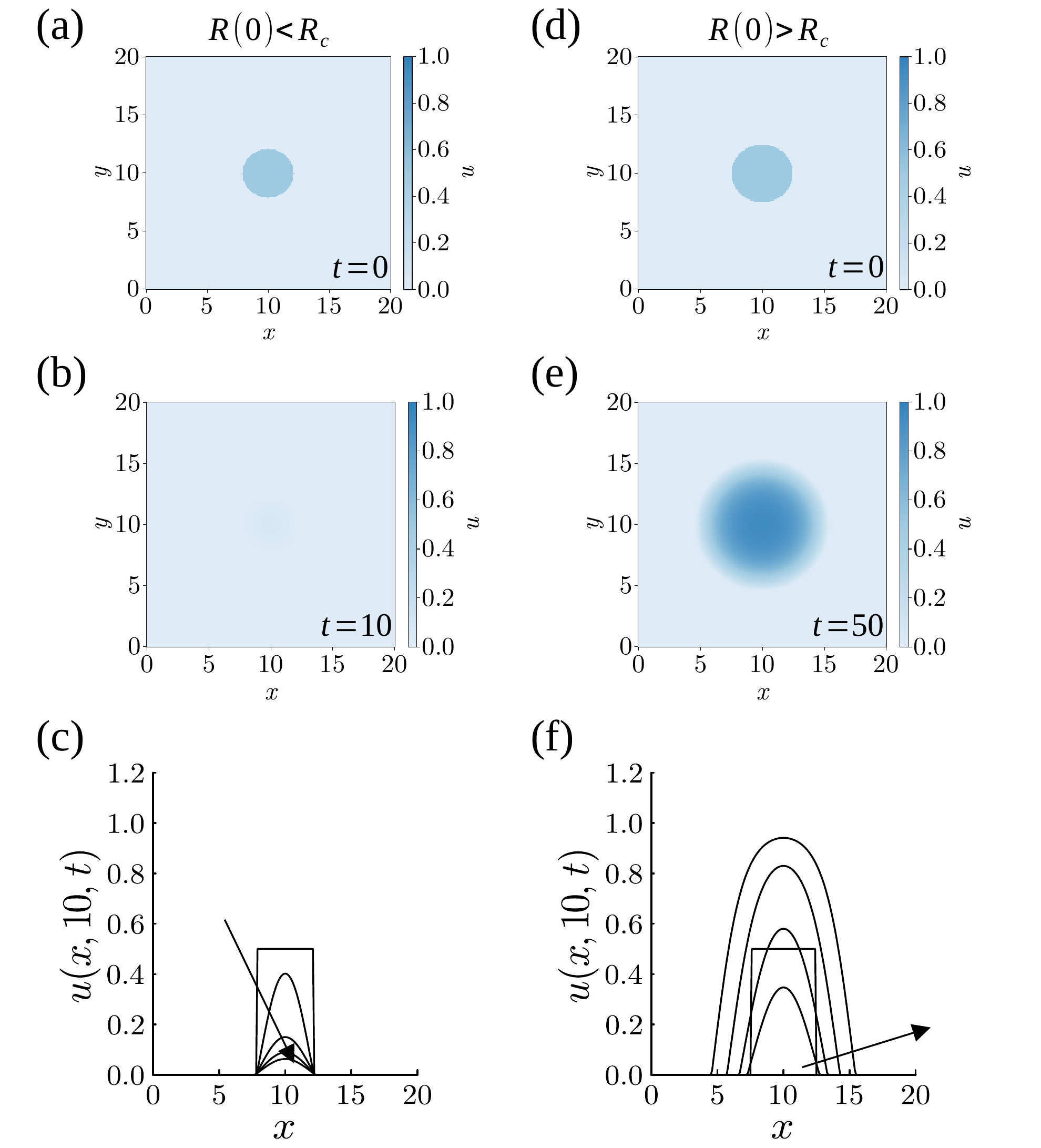}
    \caption{Numerical solutions of (\ref{eq:Level_set_model}) for a one--phase invasion problem with disc--shaped initial condition for $u(x,y,0)$.  Results in (a)--(c) correspond to setting $U(x,y) = 1/2$ within a disc with $R(0)=2.1$, whereas (d)--(f) correspond to setting $U(x,y) = 1/2$ within a disc with $R(0)=2.5$.  All results have $V(x,y)=0$ and $\kappa_u=0.2$.  (a)--(b) shows the evolution of the solution with $R(0) < R_c \approx 2.4048$ leading to extinction, whereas (d)--(e) show the evolution of the solution with $R(0) > R_c \approx 2.4048$ leading to survival.  Profiles in (c) show $u(x,10,t)$ for $R(0) < R_c$ at $t=0,1,4,7,10$, and (f) shows $u(x,10,t)$ for $R(0) > R_c$ at $t=0, 5, 20, 35, 50$. In (c) and (f) the arrows show the direction of increasing $t$.}
    \label{Figure_3}
\end{figure}

\begin{figure}
    \centering
    \includegraphics[width=1.0\textwidth]{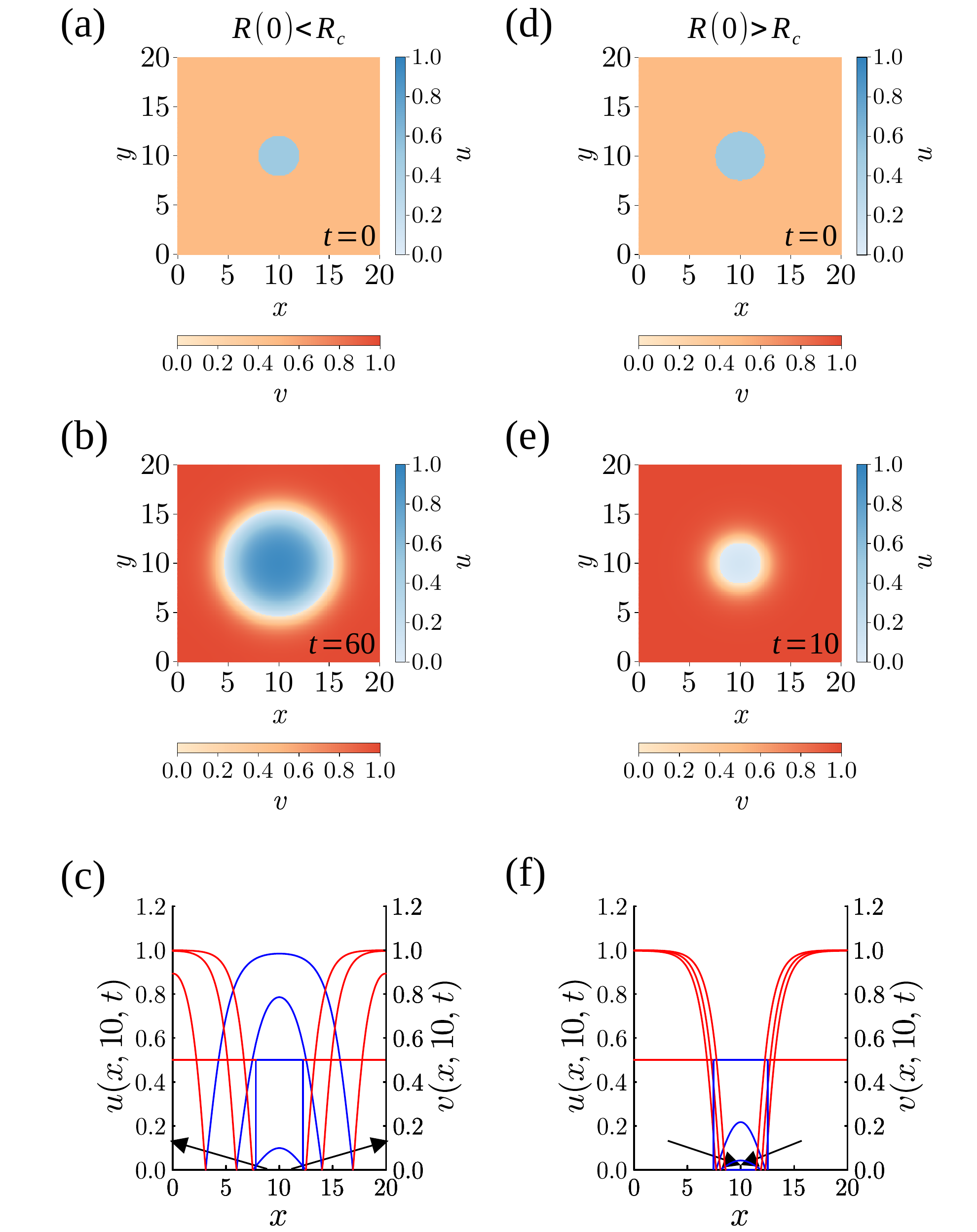}
    \caption{Numerical solutions of (\ref{eq:Level_set_model}) for a two--phase invasion problem with disc--shaped initial condition for $u(x,y,0)$.  Results in (a)--(c) correspond to setting $U(x,y) = 1/2$ within a disc with $R(0)=2.1$ and $V(x,y)=1/2$ outside of the disc, with $\kappa_u=0.2$ and $\kappa_v = -0.01$.  Results in (d)--(f) correspond to setting $U(x,y) = 1/2$ within a disc with $R(0)=2.5$ and $V(x,y)=1/2$ outside of the disc, with $\kappa_u=0.2$ and $\kappa_v = 0.1$.  (a)--(b) shows the evolution of the solution with $R(0) < R_c \approx 2.4048$ leading to survival of the $u$ population, whereas (d)--(e) show the evolution of the solution with $R(0) > R_c \approx 2.4048$ leading to extinction of the $u$ population.  Profiles in (c) show $u(x,10,t)$ and $v(x,10,t)$ for $R(0) < R_c$ at $t=0,25,50,75$, and (f) shows $u(x,10,t)$ and $v(x,10,t)$ for $R(0) > R_c$ at $t=0, 3, 6, 9$. In (c) and (f) the blue curves correspond to $u$, the red curves correspond to $v$, and the arrows show the direction of increasing $t$.}
    \label{Figure_4}
\end{figure}

Comparing results in Figure \ref{Figure_3}--\ref{Figure_4} illustrates how the well--established spreading--vanishing dichotomy for the one--phase Fisher--Stefan model, summarized above in Section \ref{sec: survival preamble}, fails to predict the outcomes of very similar two--phase problems.  Simulations in Figure \ref{Figure_3} involve a one--phase problem by setting $U(x,y) = 1/2$ inside a disc of radius $R(0)$ centered at the middle of the $20 \times 20 $ domain, and setting $V(x,y)=0$ outside $R(0)$.  For consistency with later results (that cannot be formulated with one spatial variable) we solve (\ref{eq:Level_set_model}) using our level set formulation in $\mathbb{R}^2$ in Cartesian coordinates as described in Section \ref{sec:numericalmethods}.  Figure \ref{Figure_3}(a) shows an initial condition where $R(0) < R_{\textrm{c}}$, and we see that by $t=10$ in Figure \ref{Figure_3}(b) that the population appears to be close to extinction.  Profiles in Figure \ref{Figure_3}(c) show a cross section of $u(x,y,t)$ along the horizontal line $y=10$, where we see that the initial density rapidly decays to zero.  In contrast, results in Figure \ref{Figure_3}(d)--(f) show the solution where $R(0) > R_{\textrm{c}}$ and we see that the population survives and spreads.  These results are consistent with the well--known spreading--vanishing dichotomy for the one--phase Fisher--Stefan model.

Simulations in Figure \ref{Figure_4} show results for  a closely--related two--phase problem with $U(x,y) = 1/2$ inside a disc of radius $R(0)$ centered at the middle of the $20 \times 20 $ domain, and $V(x,y)=1/2$ outside the disc.  The initial radius in Figure \ref{Figure_4}(a) is precisely the same as in Figure \ref{Figure_3}(a), and yet the $u$ population survives and spreads in the two--phase problem, which is exactly the opposite outcome of the one--phase simulation.  Similarly, $R(0)$ in Figure \ref{Figure_4}(d) is precisely the same as in Figure \ref{Figure_3}(d), and yet here in the two--phase setting the $u$ population becomes extinct, which is the opposite outcome of the one--phase scenario.

\begin{figure}
    \centering
    \includegraphics[width=1.0\textwidth]{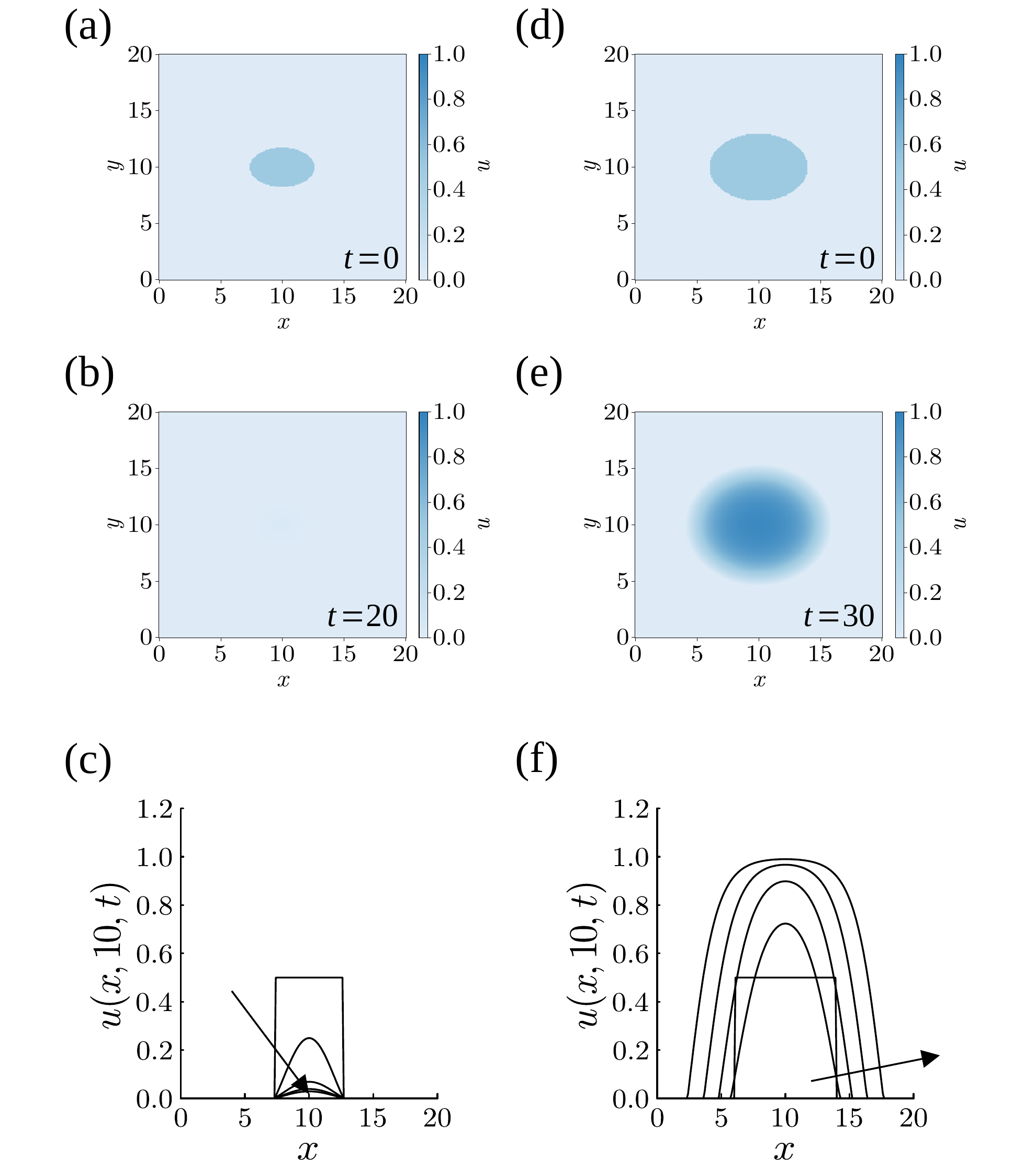}
    \caption{Numerical solutions of (\ref{eq:Level_set_model}) for a one--phase invasion problem with ellipse--shaped initial condition for $u(x,y,0)$.  Results in (a)--(c) correspond to setting $U(x,y) = 1/2$ within the ellipse where the semi--major and the semi--minor axes are of length $\sqrt{7}$  and $\sqrt{3}$, respectively.  Results in  (d)--(f) correspond to setting $U(x,y) = 1/2$ within the ellipse where the semi--major and semi--minor axes are of length is $4$ and $3$, respectively.  All results have $V(x,y)=0$ and $\kappa_u=0.2$.  (a)--(b) shows the evolution of the solution from the smaller ellipse leading to extinction, and (d)--(e) shows the evolution of the solution from the larger ellipse leading to survival.  Profiles in (c) show $u(x,10,t)$ associated with the smaller ellipse at $t=0, 2, 8, 14, 20$, and (f) shows $u(x,10,t)$ associated with the larger ellipse at $t=0, 5, 20,35, 50$. In (c) and (f) the arrows show the direction of increasing $t$.}
    \label{Figure_5}
\end{figure}

\begin{figure}
    \centering
    \includegraphics[width=1.0\textwidth]{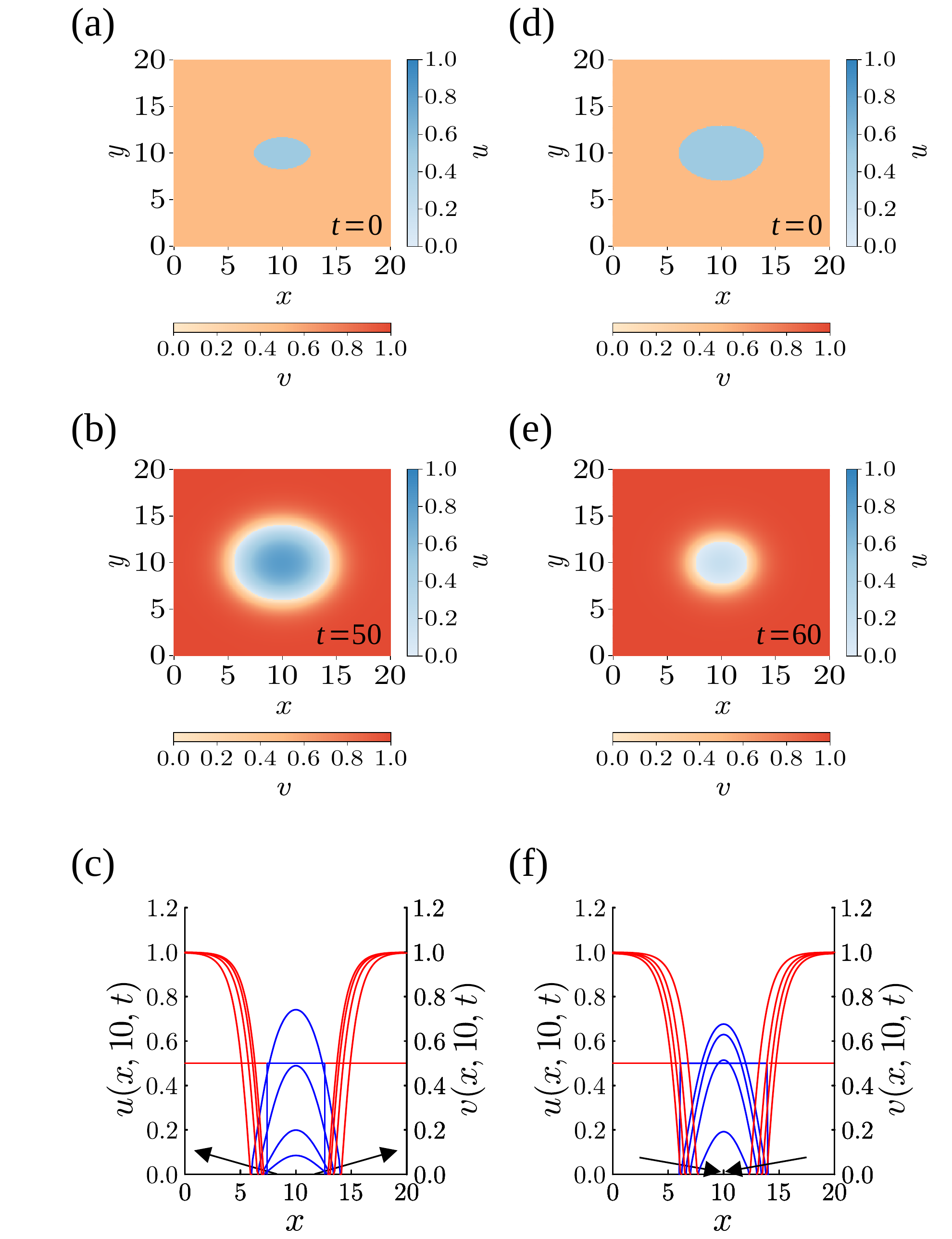}
    \caption{Numerical solutions of (\ref{eq:Level_set_model}) for a two--phase invasion problem with ellipse--shaped initial condition for $u(x,y,0)$.  Results in (a)--(c) correspond to setting $U(x,y) = 1/2$ within the ellipse where the semi--major and the semi--minor axes are of length $\sqrt{7}$  and $\sqrt{3}$, respectively, with  $\kappa_u=0.2$ and $\kappa_v = -0.01$.   Results in (d)--(f) correspond to setting $U(x,y) = 1/2$ within the ellipse where the semi--major and the semi--minor axes are of length $4$  and $3$, respectively,  with $\kappa_u=0.2$ and $\kappa_v = 0.1$. (a)--(b) shows the evolution of the solution from the smaller ellipse leading to survival of $u$, and (d)--(e) shows the evolution of the solution from the larger ellipse leading to extinction of $u$.  Profiles in (c) show $u(x,10,t)$ and $v(x,10,t)$ associated with the smaller ellipse at $t=0, 5, 20, 35, 50$, while (f) shows $u(x,10,t)$ and $v(x,10,t)$ associated with the larger ellipse at $t=0, 6, 24,62, 60$.  In (c) and (f) the blue curves correspond to $u$, the red curves correspond to $v$, and the arrows show the direction of increasing $t$.}
    \label{Figure_6}
\end{figure}

\newpage
For completeness we provide another comparison of one-- and two--phase problems in Figures \ref{Figure_5}--\ref{Figure_6} where we consider initial elliptical--shaped $u$ populations.  Figure \ref{Figure_5} show results for a one--phase problem where the initial $u$ population is placed in the centre of the $20 \times 20$ domain inside an elliptical region with the major axis parallel to the $x$--axis.  We set $U(x,y)=1/2$ inside the ellipse, and $V(x,y)=0$ outside of the ellipse.  Results in Figure \ref{Figure_5} correspond to an initial elliptical--shaped $u$ population with semi--major and semi--minor axes of $\sqrt{7}$ and $\sqrt{3}$, respectively.  Time--dependent numerical solutions, summarized in Figure \ref{Figure_5}(b)--(c), show that the population eventually becomes extinct.  In contrast, results in Figure \ref{Figure_5}(d)--(f) illustrate that a sufficiently large elliptical population, where the length of the semi--major and semi--minor axes are $4$ and $3$, respectively, leads to survival and spreading.  This observation is consistent with the spreading--vanishing dichotomy since spreading is associated with a population that occupies a sufficiently large region, and extinction is associated with a population that does not occupy a sufficiently large region.   Results in Figure \ref{Figure_6} show the evolution of two closely--related two--phase problems.  In particular, the initial placement of the $u$ population in Figure \ref{Figure_6}(a) and (d) are identical to the initial placement of the $u$ population in Figure \ref{Figure_5}(a) and (d), respectively.  The solutions of the two--phase problems in Figure \ref{Figure_6} involves setting $V(x,y)=1/2$ outside of the central elliptical region;  time--dependent solutions in Figure \ref{Figure_6}(a)--(c) illustrate that the $u$ population survives and spreads, whereas the $u$ population becomes extinct in Figure \ref{Figure_6}(d)--(f).  As with the disc--shaped initial populations in Figures \ref{Figure_3}--\ref{Figure_4}, we again see that comparing one--phase and two--phase problems can lead to precisely the opposite outcome in terms of population survival or extinction.  \cbl Of course, as in Figures \ref{Figure_3}--\ref{Figure_4}, the differences in long--term survival or extinction in Figures  \ref{Figure_5}--\ref{Figure_6} is driven by the differences in $\kappa_v$ which affects the short--time dynamics of the interface in such a way that the long--term survival or extinction in the two--phase case is completely the opposite of associated the one--phase results. \cb

\newpage

\section{Results and discussion II: Stability of travelling wave solutions}\label{sec: stability}

\subsection{Preamble: Travelling wave solutions for a two--phase Fisher Stefan model} \label{sec: stability preamble}
As outlined in the Introduction, El-Hachem et al.~\cite{ElHachem2020} studied travelling wave solutions of a one--dimensional analogue of (\ref{eq:Non-dimensional_model}), namely
\begin{align}
\label{eq:PartialDiffUNonDim}
&\frac{\partial u}{\partial t} =  \frac{\partial^{2} u}{\partial x^{2}} + u (1-u), \quad - \infty < x < s(t), \\
\label{eq:PartialDiffVNonDim}
&\frac{\partial v}{\partial t} = D \frac{\partial^{2} v}{\partial x^{2}} + \lambda v (1-v), \quad  s(t) < x < \infty,
\end{align}
where the boundary conditions are given by
\begin{align}
\label{eq:BCNeumannNonDim}
& \lim_{x \to -\infty }u(x,t) = 1,  \quad \lim_{x \to \infty }v(x,t) = 1, \\
&u(s(t),t) = v(s(t),t) = 0, \\
&\frac{\mathrm{d} s(t)}{\mathrm{d} t} = -\kappa_u  \left. \frac{\partial v}{\partial x} \right|_{x=s(t)} - \kappa_v \left. \frac{\partial u}{\partial x} \right|_{x=s(t)}. \label{eq:NondimbcDirichlet}
\end{align}
While the moving boundary model is defined on an infinite domain, El-Hachem et al.~\cite{ElHachem2020} obtained numerical solutions of this PDE model by working with an appropriately truncated domain.  In this one--dimensional model the moving boundary is defined as a point $x = s(t)$, whereas in our two--dimensional analogue the moving boundary is defined by a curve $\partial \Omega(t)$.    Motivated by preliminary numerical solutions, El-Hachem et al.~\cite{ElHachem2020} sought travelling wave solutions by re--writing the governing equations in the usual travelling wave coordinate $z = x - ct$, where $c$ is the constant travelling wave speed and seeking solutions of the form $u = \mathcal{U}(z)$ and $v= \mathcal{V}(z)$.    Using a combination of phase--plane analysis, perturbation methods and full time--dependent solutions of the PDE model, El-Hachem showed that this one--dimensional model gives rise to travelling wave solutions with $c>0$ whereby population $u$ invades into $v$ which might represent malignant invasion in the case that $u$ represents a population of cancer cells and $v$ represents a population of healthy surrounding tissue. Interestingly,  this model also gives rise to travelling wave solutions with $c < 0$ whereby population $v$ invades into population $u$ which might represent malignant retreat.  Furthermore, the same model also gives rise to stationary travelling wave solutions with $c=0$.  This intermediate case turns out to be of both mathematical and practical interest because the dynamical system governing $\mathcal{U}(z)$ and $\mathcal{V}(z)$ has exact solutions for $c=0$, and these solutions can be used to construct approximate perturbation solutions that are valid for $|c| \ll 1$~\cite{ElHachem2020}.  Overall, the one--dimensional two--phase model leads to sharp--fronted travelling wave solutions that move with any wave speed $c \in (-\infty, \infty)$, and one way to interpret this result is that this simple moving boundary model is more biological relevant than the well--studied Fisher--Kolmogorov model which fails to predict a sharp front, and only supports invading travelling wave solutions with $c \ge 2$.

One of the limitations of the study by El-Hachem et al.~\cite{ElHachem2020} is that they only considered solutions in one spatial dimension, rather than the more biologically  realistic two--dimensional setting.  In two dimensions, the evolution of the interface can be more complicated, because not only the position, but the shape of the interface can change over time.  We analyze the linear stability of planar fronts of the two--phase invasion model to transverse perturbations.  This analysis can reveal front patterns such as fingering and morphological instabilities.  Linear stability analysis has long--been of interest in the study of reaction--diffusion problems~\cite{Yang2002}, and industrial problems~\cite{Straint1988}, but the stability of solutions for two--phase biological invasion models has received less attention.

\subsection{Stability of travelling wave solutions for the two--phase invasion model}

We first generate some preliminary numerical solutions to visually explore the stability of travelling wave solutions using our level--set numerical method.  To achieve this we solve a range of problems on the rectangular domain $\mathcal{D}$ with boundary conditions summarized in Figure \ref{Figure_1}(d).  There are several approaches we could use to generate the travelling wave solutions and to incorporate the transverse perturbation.  The most straightforward approach is to fix particular values of $\kappa_u$, $\kappa_v$, $D$ and $\lambda$ and set $U(x,y) = 1$ at all locations where $x < X$, $V(x,y)= 1$ where $x > X$, and $U=V=0$ along the vertical line $x=X$.  Solving the system (\ref{eq:Non-dimensional_model}) for this initial condition leads to long--time travelling waves provided that $\mathcal{D}$ is sufficiently wide.  The speed of the travelling wave will depend upon the choices of $\kappa_u$, $\kappa_v$, $D$ and $\lambda$, and can be estimated from the long--time PDE solutions~\cite{ElHachem2020}.   Generating travelling wave solutions in this way allows us to check the accuracy of our level--set numerical method by comparing the shape of these long--time travelling wave solutions predicted by the level--set numerical method with various approximate perturbation solutions reported by El-Hachem et al.~\cite{ElHachem2020} (not shown).  Since travelling wave solutions are translationally invariant, we can shift the established travelling wave profile so that the vertical moving boundary is located half way along $\mathcal{D}$.  At this point a small transverse sinusoidal perturbation with wavenumber $q$ can be added, and the resulting profiles can be used as initial conditions to solve the system (\ref{eq:Non-dimensional_model}) to explore whether the amplitude of the transverse perturbation grows or decays.  There are also other options for specifying the travelling wave solution with a transverse perturbation, and we will discuss a different approach later in this section.

\begin{figure}
\centering
\includegraphics[width=1.0\textwidth]{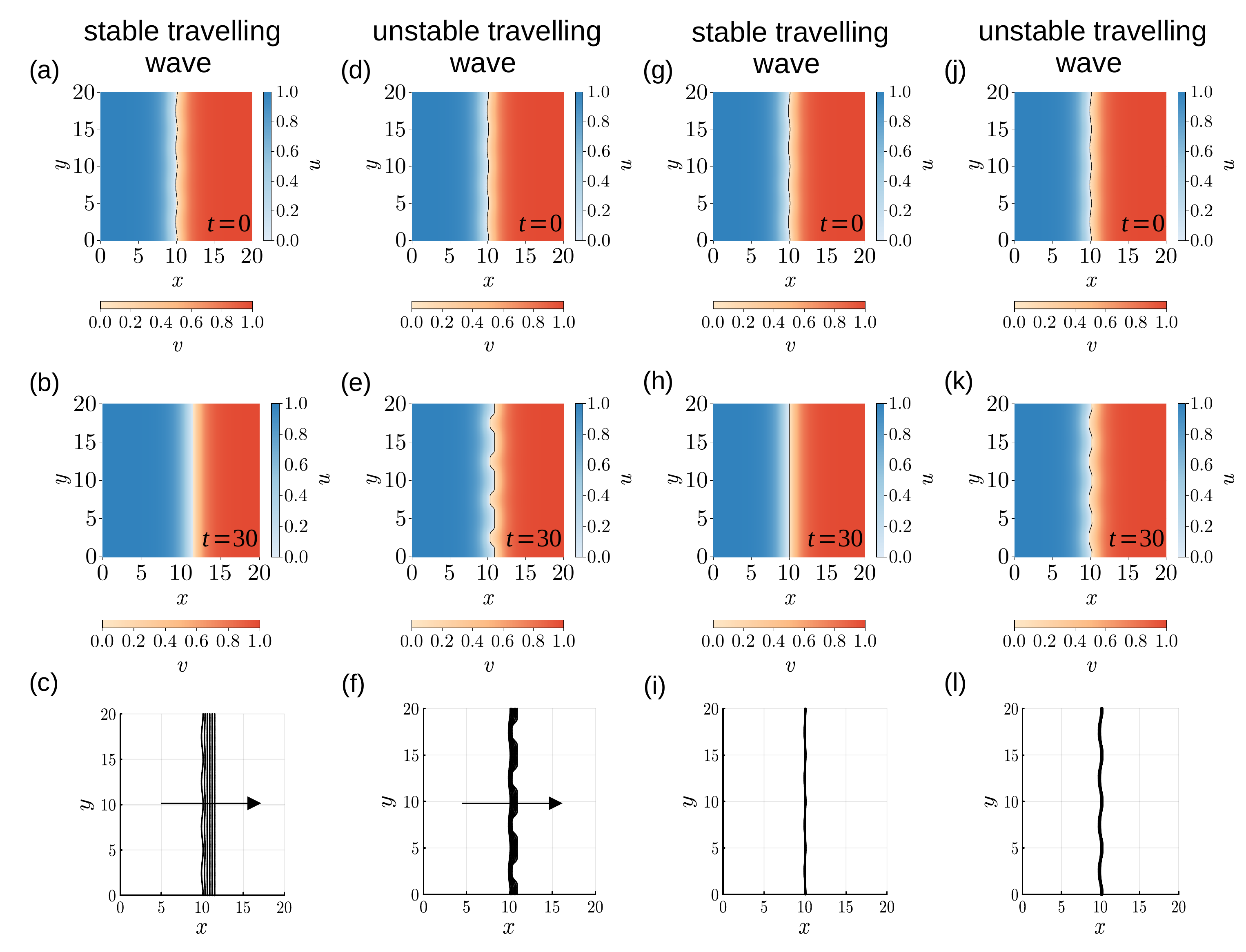}
\caption{Preliminary numerical exploration of travelling wave stability with $D = \lambda = 1$.  Results in the top row show various travelling wave solutions on a $20 \times 20$ domain where a transverse perturbation has been added.  Profiles in the top row are used as initial conditions to solve (\ref{eq:Level_set_model}).  Solutions at $t=30$ are given in the middle row, and the dynamics of the interface movement is given lower row at $t = 0, 5, 10, 15, 20, 25, 30$. Arrows in (c) and (f) show the direction that the travelling wave solution is moving, note that arrows are not included in (i) and (l) because the travelling wave solutions in these cases are stationary. All results have $q = 2 \pi / 5$. (a)--(c) $\kappa_u=0.2$, $\kappa_v=0.1$ and $c=0.05$.  (d)--(f) $\kappa_u=-0.05$, $\kappa_v=-0.1$ and  $c=0.027$. (g)--(i) $\kappa_u = \kappa_v=0.1$ and $c=0$. (j)--(l) $\kappa_u=\kappa_v=-0.1$ and  $c=0$. }
\label{Figure_7}
\end{figure}

The profile in Figure \ref{Figure_7}(a) shows a perturbed travelling wave that corresponds to $D = \lambda = 1$,  $\kappa_u=0.2$, $\kappa_v=0.1$ and $c=0.05$, and the perturbation wavenumber is $q=2\pi/5$.  The numerical solution at $t=30$ in Figure \ref{Figure_7}(b) suggests that the travelling wave solution is stable in the sense that the perturbation introduced at $t=0$ decays with time so that the moving front at $t=30$ appears to be a straight vertical line.  The dynamics of this problem are summarized in Figure \ref{Figure_7}(c) where we show the location of the interface at $t=0, 5, 10, 15, 20, 25, 30$.  We see that the amplitude of the perturbations decays with time as the front continues to move in the positive $x$--direction.  In contrast, Figure \ref{Figure_7}(d) shows a perturbed travelling wave corresponding to $D = \lambda = 1$,  $\kappa_u=-0.05$, $\kappa_v=-0.1$ and $c=0.027$, and again the perturbation wavenumber is $q=2\pi/5$.  The numerical solution at $t=30$ in Figure \ref{Figure_7}(e) indicates that the travelling wave solution is unstable in the sense that the amplitude of the perturbation grows with time such that the travelling front at $t=30$ has evolved to have a pronounced nonlinear scallop--shaped front.  The dynamics of the evolution of the interface are summarized in Figure \ref{Figure_7}(f).  Additional results in Figure \ref{Figure_6}(g)--(i) and Figure \ref{Figure_6}(j)--(l) show two further examples, one apparently stable and another apparently unstable, respectively.  In these final two cases we set $\kappa_u = \kappa_v$ with $D = \lambda = 1$ so that the travelling waves are stationary and $c=0$.

It is interesting to compare these results in Figure \ref{Figure_7} with the previous results reported by Tam and Simpson~\cite{Tam2023} who studied the stability of two--dimensional fronts in a one--phase Fisher--Stefan invasion model.  In that previous work, Tam and Simpson showed that travelling wave solutions with $c>0$ were linearly stable to transverse perturbations, whereas travelling wave solutions with $c < 0$ were linearly instable.  One simple interpretation of those previous results was to consider travelling waves where $u$ advances in the positive $x$ direction with $\partial u / \partial x < 0$ at the front are linearly stable, whereas travelling waves where $u$ moves in the negative $x$ direction with  $\partial u / \partial x < 0$ at the front as linearly unstable.  While results in Figure \ref{Figure_7}(a)--(c) are consistent with these previous one--phase results, the simulation results in Figure \ref{Figure_7}(d)--(f) suggest that the opposite outcome can occur in a two--phase extension, since here we have $u$ advancing in the positive $x$ direction with $\partial u / \partial x < 0$ at the front appearing to be unstable to transverse perturbations.

To provide mathematical insight into the numerical explorations in Figure \ref{Figure_7} we use linear stability analysis to provide mathematical insight into the stability properties of travelling wave solutions of the two--phase Fisher--Stefan model.  We follow the approach of M\"uller and van Saarloos~\cite{Muller2002}, and Tam and Simpson~\cite{Tam2023} by expanding the travelling wave front position $x=L(y,t)$ as
\begin{equation}
  \label{eq:results_perturbation_front}
  L(y,t) = ct + \varepsilon \exp ({\mathrm{i} qy + \omega t}) + \mathcal{O}(\varepsilon^2),
\end{equation}
where $\varepsilon \ll 1$ is a measure of the amplitude of the perturbation, $q$ is the wavenumber and $\omega$ is the growth rate.  Our approach involves numerically solving a boundary value problem that can be thought of as an eigenvalue problem for the real component of the growth rate, $\omega$.  Upon solving the boundary value problem, we obtain a dispersion relation that describes the relationship between $\omega$ and $q$.   A perturbation with a wavenumber $q$ is linearly stable if $\omega(q) < 0$, or linearly unstable if $\omega(q) > 0$.

To proceed, we write
\begin{equation}
  \label{eq:variable_z}%
  \xi = x - ct - \varepsilon\exp(\mathrm{i} qy + \omega t),
\end{equation}
and to be consistent we also perturb the signed distance function, $\phi = x - ct -  \varepsilon\exp(\mathrm{i} qy + \omega t)$.  Expanding $u(x,y,t)$ and $v(x,y,t)$ gives
\begin{subequations}
\label{eq:perturbation_density_u_v}
\begin{gather}
u(x,y,t) = u_0(\xi) + \varepsilon u_1(\xi)\exp(\mathrm{i} qy + \omega t) + \mathcal{O}(\varepsilon^2), \label{eq:results_perturbation_density_u}\\
v(x,y,t) = v_0(\xi)  + \varepsilon v_1(\xi)\exp(\mathrm{i} qy + \omega t) + \mathcal{O}(\varepsilon^2).  \label{eq:results_perturbation_density_v}%
\end{gather}
\end{subequations}
Implementing this change of variables and substituting the expansions for $u(x,y,t)$ and $v(x,y,t)$ into the level--set formulation shows that, as expected, the leading order problem for $u_0(\xi)$ and $v_0(\xi)$ is the same travelling wave problem for $\mathcal{U}(z)$ and $\mathcal{V}(z)$ studied by El-Hachem et al.~\cite{ElHachem2020}
\begin{subequations}
    \label{eq:leading_order_model}
    \begin{gather}
        \frac{\textrm{d}^2u_0}{\textrm{d}\xi^2} + c\frac{\textrm{d}u_0}{\textrm{d}\xi} + u_0\left(1-u_0\right) = 0 \quad \mathrm{ on } \quad  -\infty< \xi < 0, \label{eq:leading_order_u0}\\
        D\frac{\textrm{d}^2v_0}{\textrm{d}\xi^2} + c\frac{\textrm{d}v_0}{\textrm{d}\xi} + \lambda v_0\left(1-v_0\right) = 0 \quad \mathrm{ on } \quad 0< \xi <\infty,  \label{eq:leading_order_v0}\\
       \lim_{\xi \to -\infty} u_0(\xi) =\lim_{ \xi\to \infty} v_0(\xi) = 1, \label{eq:leading_order_boubdary_contions} \\
        u_0(0) = \;v_0(0) = 0,  \label{eq:leading_order_initial_contions}\\
        c = -\kappa_u\;\frac{\textrm{d}u_0(0)}{\textrm{d}\xi} - \kappa_v\;\frac{\textrm{d}v_0(0)}{\textrm{d}\xi},\label{eq:leading_order_velocity}
    \end{gather}
\end{subequations}
We analyse this travelling wave problem for $u_0$ and $v_0$ using two  different approaches.  First, we take a numerical approach by truncating the infinite domain, discretizing the derivative terms on a uniform meshing of the truncated domain, and then solve the resulting nonlinear algebraic system of equations in the context of a shooting method by treating $c$ as an unknown parameter to be determined as part of the numerical solution.  Second, we follow the approach of El-Hachem et al.~\cite{ElHachem2020}, noting we can solve the system (\ref{eq:leading_order_model}) in the special case that $c=0$, and we use this exact solution to construct a perturbation solution for $u_0$ and $v_0$ for $|c| \ll 1$.  A detailed description of both approaches is given in the Appendix, where we show that results obtained using shooting method numerical approach are visually indistinguishable from the perturbation solutions.

The $\mathcal{O}(\varepsilon)$ problem governing the correction terms $u_1$ and $v_1$ is
\begin{subequations}
    \label{eq:first_order_model}
    \begin{gather}
        \frac{\textrm{d}^2u_1}{\textrm{d}\xi^2} + c\frac{\textrm{d}u_1}{\textrm{d}\xi} + [1-\omega-q^2-2\;u_0]u_1 +(\omega+q^2)\;  \frac{\textrm{d}u_0}{\textrm{d}\xi}=0 \quad \mathrm{ on } \quad  -\infty< \xi < 0, \label{eq:first_order_u1}\\
        D\frac{\textrm{d}^2v_1}{\textrm{d}\xi^2} + c\frac{\textrm{d}v_1}{\textrm{d}\xi} + [\lambda-\omega-Dq^2-2\lambda\;v_0]v_1 +(\omega+Dq^2)\;  \frac{\textrm{d}v_0}{\textrm{d}\xi}=0\quad \mathrm{ on } \quad 0< \xi <\infty,  \label{eq:first_order_v1}\\
       \lim_{\xi \to -\infty} u_1(\xi) =\lim_{\xi \to \infty} v_1(\xi) = 0, \label{eq:first_order_boundary_conditions} \\
        u_1(0) = \; v_1(0) = 0,  \label{eq:first_order_initial_conditions}\\
        \omega = -\kappa_u\;\frac{\textrm{d}u_1(0)}{\textrm{d}\xi} - \kappa_v\;\frac{\textrm{d}v_1(0)}{\textrm{d}\xi}.\label{eq:first_order_growth_rate}
    \end{gather}
\end{subequations}
Similar to (\ref{eq:leading_order_model}), we truncate the infinite domain, discretize the derivative terms on a uniform discretization of the truncated domain, and then solve the resulting  algebraic system of equations in the context of a shooting method where we use the value of $c$ from (\ref{eq:leading_order_model}) and we treat the growth rate $\omega$ as an unknown parameter.  Unlike (\ref{eq:leading_order_model}) we have not found any exact or approximate  perturbation solutions of (\ref{eq:first_order_model}).  Therefore we focus on solving (\ref{eq:first_order_model}) numerically using a shooting method that is described in the Appendix.  To obtain the dispersion relationship, $\omega(q)$, we repeatedly solve (\ref{eq:leading_order_model})--(\ref{eq:first_order_model}) treating $D$, $\lambda$, $\kappa_u$ and $\kappa_v$ as inputs into the numerical procedure, which then provides estimates of $u_0$, $v_0$, $u_1$, $v_1$, $c$ and $\omega$ as outputs.  Repeating this process for various choices of $q$ provides estimates of $\omega(q)$. A travelling wave solution is considered to be linearly stable if the dispersion relation satisfies $\omega(q) < 0$ for all wave numbers $q$, while it is considered linearly unstable if $\omega(q) > 0$ for any wave number $q$.

Before presenting our results it is worth noting that our perturbation solution gives us another method to generate initial conditions for the level-set method to study the stability of travelling waves with transverse perturbations.  Given numerical estimates of $u_0$, $u_1$, $v_0$ and $v_1$ we can specify a perturbed initial condition for the level-set method as
\begin{subequations} \label{eq:perturbedic}
\begin{align}
u(x,y,0) &= u_0\left(x - \beta - \varepsilon\cos(qy)\right) + \varepsilon u_1\left(x - \beta - \varepsilon\cos(qy)\right)\cos(qy),\label{eq:results_numerical_ic_u}\\
v(x,y,0) &= v_0\left(x - \beta - \varepsilon\cos(qy)\right) + \varepsilon v_1\left(x - \beta - \varepsilon\cos(qy)\right)\cos(qy),\label{eq:results_numerical_ic_V}\\
\phi(x,y,0) &= x - \beta - \varepsilon\cos(qy),\label{eq:results_numerical_ic_phi}
\end{align}
\end{subequations}
\cbl where $\beta$ is a translation in $x$. \cb In all numerical solutions we choose $\beta$ to be sufficiently-large so that these solutions are consistent with the boundary conditions, and we set $\varepsilon = 0.1$ in (\ref{eq:perturbedic}) for all time--dependent PDE simulations presented in this study, including the preliminary results in Figure \ref{Figure_7}.

To provide an additional check of our linear stability results we independently estimate the growth/decay rate using short--time numerical solutions of the full time--dependent PDE model, (\ref{eq:Non-dimensional_model}).  At each time step of the numerical solution we estimate the amplitude of the perturbation
\begin{subequations}
\begin{align}
A(t) &= \dfrac{X_{\textrm{max}}(t) - X_{\textrm{min}}(t)}{2}, \text{where},\\
X_{\textrm{max}}(t) &= \max_{y}[x  \mid \phi(x,y,t)=0], \\
X_{\textrm{min}}(t) &= \min_{y}[x \mid \phi(x,y,t)=0].
\end{align}
\end{subequations}
To compute these quantities we use linear interpolation to find positions $x_j$ such that $\phi(x,y_j,t_k)=0$ for the $j$th row in the discretization of $\mathcal{D}$, at time $t=t_k$.  The maximum and minimum values in the set $\{x_j\}$ determine $X_{\textrm{max}}(t)$ and $X_{\textrm{min}}(t)$, respectively.  To estimate $\omega$ we follow Equation (\ref{eq:results_perturbation_front}) and assume that the perturbation amplitude grows or decays exponentially with time, which is reasonable over a sufficiently short time interval. Using the Polynomials.jl package in Julia we obtain a least--squares fit of $\log(A) = \omega_{\textrm{n}} t + C$, where $C$ is a constant and $\omega_{\textrm{n}}$ is the growth rate estimated from the numerical solution of the full time--dependent PDE problem, (\ref{eq:Non-dimensional_model}).  We use the interval $t \in [0.1,1]$ to compute $\omega_{\textrm{n}}$, and in all cases we plot $\log(A)$ as a function of time to provide a visual check that the relationship is well described by a linear function over this time interval.

\begin{figure}
\centering
\includegraphics[width=1.00\textwidth]{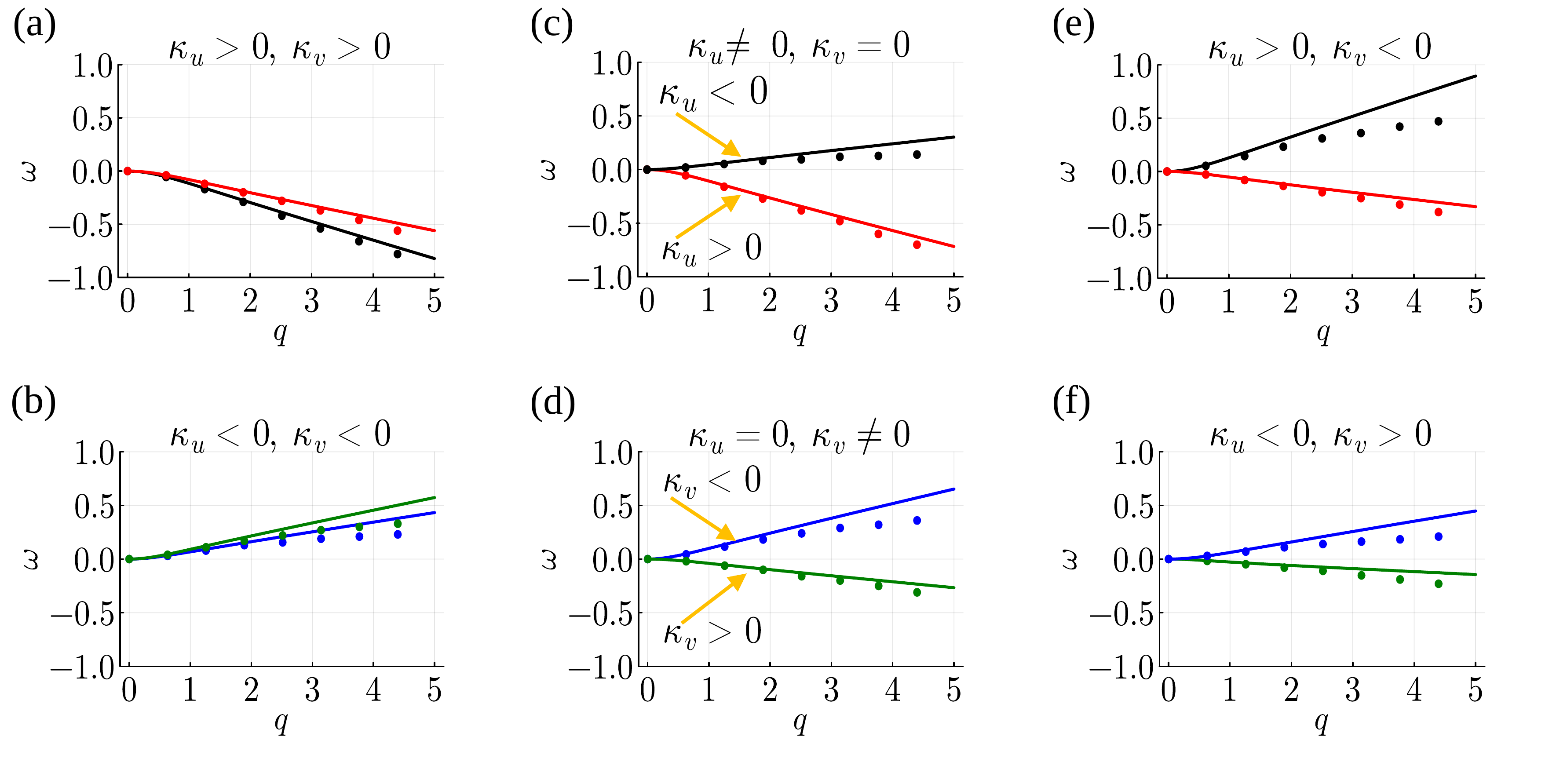}
\caption{Comparison of the dispersion relationship $\omega(q)$ obtained from the linear stability analysis (solid lines) and using full time--dependent solutions of (\ref{eq:Level_set_model}) (dots).  Time--dependent numerical solutions of the PDE model are obtained on a $30 \times 30$ domain. (a) $\kappa_u=0.2$, $\kappa_v=0.1$ (black), and $\kappa_u=0.1$, $\kappa_v=0.1$  (red). (b) $\kappa_u=-0.05$, $\kappa_v=-0.1$ (blue), and $\kappa_u=-0.1$, $\kappa_v=-0.1$  (green). (c) $\kappa_u=0.3$, $\kappa_v=0$ (red) and $\kappa_u=-0.1$, $\kappa_v=0$ (black). (d) $\kappa_u=0.0$, $\kappa_v=0.1$ (green), and $\kappa_u=0$, $\kappa_v=-0.2$  (blue). (e) $\kappa_u=0.2$, $\kappa_v=-0.05$ (red), and $\kappa_u=0.1$, $\kappa_v=-0.3$  (black). (f) $\kappa_u=-0.1$, $\kappa_v=0.2$  (green), and $\kappa_u=-0.2$, $\kappa_v=0.1$ (blue).}
\label{Figure_8}
\end{figure}

Using this procedure we compare estimates of $\omega_{\textrm{n}}(q)$ and $\omega(q)$ obtained by linear stability analysis in Figure \ref{Figure_8} for a range of $\kappa_u$ and $\kappa_v$ values.  Results in Figure \ref{Figure_8}(a)--(b) correspond to the case in which $\kappa_u$ and $\kappa_v$ have the same sign;  we observe that when both constants are positive the travelling wave solutions are linearly stable, whereas when both constants are negative the travelling wave solutions are linearly unstable.  Results in Figure \ref{Figure_8}(c)--(d) correspond to cases for which either $\kappa_u=0$ or $\kappa_v=0$, where we see that some parameter combinations lead to linearly stable travelling wave solutions while others lead to linearly unstable travelling wave solutions.  Figure \ref{Figure_8}(e)--(f) summarize results where $\kappa_u$ and $\kappa_v$ are different sign, and again we obtain a suite of mixed results in terms of linear stability.  Comparing our estimates of $\omega$ with $\omega_{\textrm{n}}$ over the 96 different combinations of $\kappa_u$ and $\kappa_v$ in Figure \ref{Figure_8} shows that both approaches compare well across the broad range of conditions.  Some results, particularly for perturbations with larger $q$, suggest that the growth rate predicted by the linear stability analysis slightly overestimates $\omega_{\textrm{n}}$.  There are many potential reasons for this discrepancy, perhaps the most obvious is that all results in Figure \ref{Figure_8} correspond to the same finite difference mesh, but we anticipate that time--dependent PDE solutions with larger $q$ may require a finer mesh to accurately resolve the initial perturbation.

\newpage

A further comment about the results in Figure \ref{Figure_8} relates to the apparent linear increase of $\omega$ and $\omega_{\textrm{n}}$ as a function of wavenumber $q$ for the linearly unstable cases. Such behaviour is indicative of a type of catastrophic instability, with increasingly larger wavenumbers (smaller wavelengths) becoming ever more unstable, that is associated with the problem being ill--posed.   Such ill--posedness makes it difficult to run numerical simulations, since truncation errors that develop on a finer mesh may become more problematic than those that occur on a coarser mesh.

\begin{figure}
\centering
\includegraphics[width=1.0\textwidth,height=0.75\textheight]{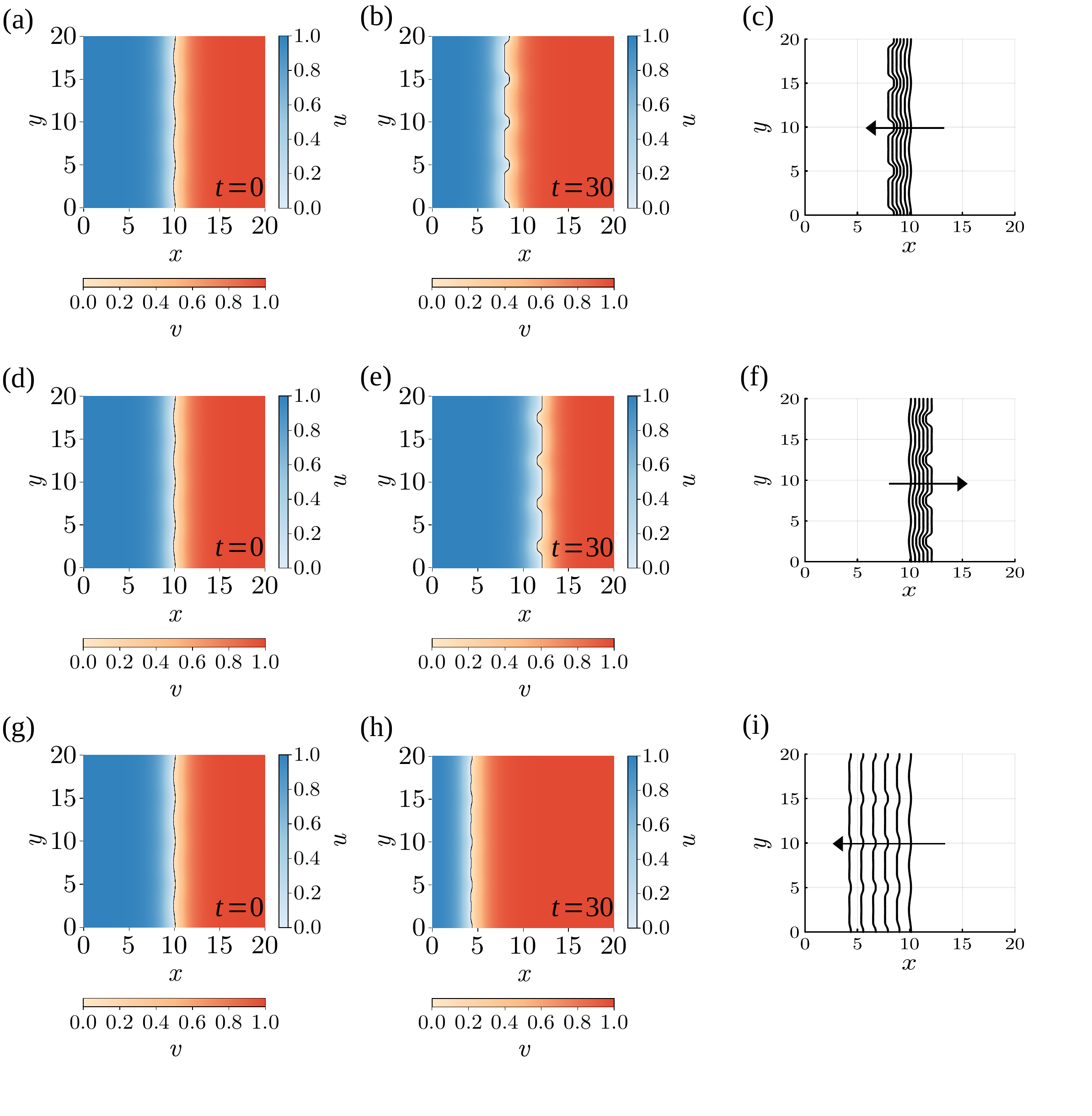}
\caption{Long-time numerical exploration of travelling wave stability with $D = \lambda = 1$, and perturbations with wavenumber $q=2\pi/5$.  The left--most column shows the perturbed initial condition, the central column shows the solution of the time--dependent PDE model at $t=30$ and the right--most column summarises the dynamics of the interface at $t=0, 5,10, 15, 20, 25, 30$.  (a)--(c) $\kappa_u=-0.1$, $\kappa_v=0$ and $c=-0.056$. (d)--(f)  $\kappa_u=0$, $\kappa_v=-0.1$, $c=0.056$. (g)--(i) $\kappa_u=-0.2$, $\kappa_v=0.1$, $c=-0.19$. }\label{Figure_9}
\end{figure}

Results in Figure \ref{Figure_8} illustrate that our two approaches to examine the linear stability of various travelling wave solutions provide consistent results by examining the short--time growth and decay rates of the amplitudes of transverse perturbations.  One of the limitations of the linear stability analysis is that it is inherently focused on short--time behaviour since we make the implicit assumption that perturbation amplitudes are small.  For those travelling waves that are linearly unstable it is unclear how these perturbations grow over longer time scales.  Figure \ref{Figure_9} illustrates the evolution of some linearly unstable travelling wave solutions over the interval $0 < t < 30$ whereas the numerical estimates of the growth rates in Figure \ref{Figure_8} corresponded to $0.1 < t < 1$ to capture the short--time exponential growth or decay in amplitude. Results in Figure \ref{Figure_9}(a)--(c) show the evolution of a linearly unstable travelling wave with $c < 0$ where we see the scallop--shaped front moving in the negative $x$--direction, whereas results in Figure \ref{Figure_9}(d)--(f) show the evolution of a linearly unstable travelling wave with $c > 0$ where the scallop--shaped front moves in the positive $x$--direction.  In both cases we see that the amplitude of the initial perturbation grows over a short time period and then stops growing over a longer timescale.  Results in Figure \ref{Figure_9}(g)--(i)  show the evolution of a linearly unstable travelling wave with $c < 0$ and again we see that the initial amplitude grows with time, and then the scallop--shaped front moves in the negative $x$--direction without further growth in amplitude of the perturbations.    Despite all three cases shown in Figure \ref{Figure_9} corresponding to linearly unstable cases, we do not see continual growth in the amplitude of the perturbations, instead we see that the growth appears to saturate at some finite amplitude~\cite{Tam2023}. Presumably the growth in amplitude slows as a result of some feature that is not accounted for in the linear stability analysis, such as numerical regularisation that is known to occur with the level--set method~\cite{Chen1997,Harabetian1998}.

\newpage
As illustrated in Figures \ref{Figure_8}--\ref{Figure_9}, some choices of $\kappa_u$ and $\kappa_v$ lead to travelling wave solutions that are linearly unstable to transverse perturbations.  In the heat transfer context, linearly unstable front propagation is often studied by regularisation~\cite{Chadam1983,Yang2002,Straint1988}, where the aim can be to include additional terms in the mathematical model that can have the dual role of improving the mathematical model by incorporating additional important mechanisms, as well as stabilising modes of perturbation with large wavenumbers.  One of the disadvantages of regularisation methods is that the choice of regularisation term is not unique~\cite{Tam2023}.  In the heat transfer and fluid mechanics literature, a common approach for Stefan--type moving boundary problems is to include terms describing surface tension effects.  This is  a popular approach as incorporating surface tension leads to a mathematically tractable model, and surface tension is known to regularise otherwise ill--posed problems~\cite{Chen1997,Sethian1992,Morrow2019}.  For example, Chadam and Ortoleva~\cite{Chadam1983} show that planar melting in the classical Stefan problem is linearly stable to transverse perturbations of all wavenumbers. Conversely, they show that planar solidification of a supercooled liquid is linearly unstable, but becomes stable for sufficiently large wavenumbers if the moving boundary condition is modified to include surface tension~\cite{Chadam1983}.  Introducing surface tension involves replacing the usual interface condition $u(s(t),t)=0$ with $u(s(t),t)=\gamma \mathcal{K}$, where $\gamma > 0$ is the surface tension coefficient, and $\mathcal{K}$ is local signed curvature of the moving front.  Here we see some critical differences between using a Stefan--like moving boundary condition in the context of heat transfer and fluid mechanics  and in population biology problems.  For heat transfer and fluid mechanics problems, the moving boundary condition $u(s(t),t)=0$ is typically applied in such a way that the scaled melting temperature or scaled reference pressure is zero.  Incorporating surface tension gives a modified moving boundary condition, $u(s(t),t)=\gamma \mathcal{K}$;  if $\gamma > 0$ then we have $u > 0$ at the moving font for positive curvature, and $u < 0$ at the moving front for negative curvature.  These possibilities are physically reasonable since the temperature or pressure at the moving front can either be less than or greater than the melting temperature or reference pressure without any physical restriction.  The situation is very different in the context of mathematical models describing population invasion, where setting $u(s(t),t)=\gamma \mathcal{K}$ is unsatisfactory because population density is, by definition, non--negative $u \ge 0$.   Alternatively, another potential regularisation of Stefan--type moving boundary problems is via a kinetic term~\cite{Sethian1992,King2005}, which leads to $u =  \mu \mathcal{v}_n$ on $\partial \Omega(t)$ where $\mu$ is a kinetic parameter. But again this leads to $u > 0$ or $u < 0$, depending on the direction the interface moves, which is unsatisfactory from a population biology point of view.  Therefore, instead of applying a speculative regularisation term to our two--phase invasion model, we note that our longer term numerical simulations in Figure \ref{Figure_9} show that the amplitude of the front in linearly unstable problems does not continue to grow unbounded.  One potential reason for this is that our level--set method acts to regularize the linearly unstable fronts.

\section{Conclusions and Outlook}\label{Sec:Conclusions}
In this work we present a two--dimensional mathematical model of biological invasion describing the interaction of two populations with densities $u$ and $v$.  Each population undergoes linear diffusion and logistic growth, and the interface between the populations where $u=v=0$ moves according to a Stefan--like moving boundary condition.  This mathematical modelling framework can be used to describe biological invasion problems that involve the motion of a sharp front between the two populations, such as the invasion or retreat of a malignant population of cells surrounded by normal tissues as in Figure \ref{Figure_1}(a)--(b).  Our work significantly extends the insights provided by El-Hachem et al.~\cite{ElHachem2020} who studied travelling wave solutions in a one--dimensional Cartesian coordinate system.  Since El-Hachem et al.~\cite{ElHachem2020} restricted their focus to one--dimensional problems, they solved the mathematical model numerically using a very effective, but simple boundary fixing transformation.  The current work significantly extends our understanding of this mathematical model by employing a more sophisticated level--set method enabling us to examine biological invasion in a more general scenario where we can model both the position and shape of the moving boundary.

We use the level--set numerical method to provide insight into two different types of problems.  First, we consider various initial conditions where population $u$ is surrounded by population $v$ and we generate numerical results to investigate the long--time survival or extinction of $u$.  This problem, known previously as the spreading--vanishing dichotomy in one--phase models, is very--well studied.  In the one--phase case, the long--time survival of $u$ requires that the $u$ population must occupy a sufficiently large region.  While the mathematical conditions that govern the long--term survival or extinction of $u$ are the same regardless of whether we consider a one-- or two--phase problem, our study shows that early to intermediate time interactions between $u$ and $v$  can play out such that previously--established one--phase results turn out to be misleading in the more realistic two--phase case.  If, for example, we consider restricting $u$ to occupy a disc in a one--phase model, it is well--known that if $R(0) > R_{\textrm{c}}$ the population will survive, whereas if $R(t)$ never exceeds $R_{\textrm{c}}$ the population will eventually become extinct.  Here, we show that the two--phase moving boundary condition can impact the dynamics of the solution so that precisely the opposite outcome occurs across a range of different problems with different shaped regions containing $u$.  Therefore, we conclude that great care should be taken when using a simple one--phase models to predict long term survival or extinction~\cite{Lewis1993,Li2022} because these predictions can be very sensitive when second phase is included in the mathematical model.

The second problem we consider is to examine whether planar travelling waves in the two--phase model are stable to transverse perturbations. Again, these results significantly extend the work of El-Hachem~\cite{ElHachem2020} who considered travelling wave solutions in a one--dimensional model.  As a result, El-Hachem et al.~\cite{ElHachem2020} were unable to study whether these travelling wave solutions were stable to transverse perturbations because they did not use appropriate numerical tools to explore two--dimensional problems.  Here we use a level--set numerical method to study two--dimensional travelling waves solutions subjected to a transverse perturbation, and we examine whether the amplitude of that small perturbation grows or decays with time.  Preliminary numerical explorations indicate that some travelling wave solutions are linearly stable, while others are linearly unstable, and the difference is related to the choice of the moving boundary parameters $\kappa_u$ and $\kappa_v$.  Linear stability analysis provides estimates of the dispersion relationship $\omega(q)$, which we show to be quantitatively consistent with results from our full time--dependent PDE solutions.  For those travelling wave problems that are linearly unstable we also generate longer--time numerical solutions of the PDE model suggesting that the amplitude of the perturbation do not grow indefinitely which we attribute to numerical regularisation that is known to occur with level--set implementations.  \cbl All calculations here indicate that $\omega$ is real. \cb

While this work provides insight into a novel mathematical model of biological invasion, our results lay the foundation for several points of future investigation.  From a practical point of view, the biological mechanisms encoded in the mathematical model are relatively simple, namely linear diffusion and logistic growth.  \cbl Another point of practical interest is that all initial conditions we consider in this work involve simply connected populations.  It is relatively straightforward to take the full time--dependent PDE model and associated numerical software available on \href{https://github.com/alex-tam/TwoPhaseInvasion}{GitHub} and change the initial condition and  include additional mechanisms. \cb  Of interest would be to extend the mathematical model to incorporate different migration mechanisms such as nonlinear diffusion~\cite{McCue2019,Sherratt1990,Tam2018} or chemotaxis~\cite{Murray2002}.  Understanding how such additional mechanisms impact the spreading--vanishing dichotomy in this two--dimensional implementation of the two--phase invasion model would be very interesting, as would understanding how different migration mechanisms impact the stability of planar travelling waves.  Our work also motivates opportunities for new theoretical work, with the most obvious candidate being the development of appropriate regularisation methods for analysing linearly unstable front propagation where the moving boundary condition is written in terms of a  population density that remains non--negative.

\vspace{1cm}
\noindent
\textit{Acknowledgements} MJS and AKYT are supported by the Australian Research Council (DP200100177, DP230100406). \cbl We thank Dr Andrew Krause and an anonymous referee for helpful comments on this work. \cb

\newpage

\section*{Appendix: Solutions of boundary value problems for the linear stability analysis}\label{Numerical_BVP}
As explained in the main document, we use both a numerical and an approximate perturbation method to obtain solutions of Equation (\ref{eq:leading_order_model}), and we will now give details of both methods.  In the main document we also explained that $\xi \to z$ as $\varepsilon \to 0$, so to be consistent with El-Hachem et al.~\cite{ElHachem2020} we write all boundary value problems in terms of $z$ as the independent variable.

\subsection*{Numerical solution of the leading order boundary value problem}
The numerical solution of Equation (\ref{eq:leading_order_model}) is obtained by truncating the infinite domain and considering the boundary value problem on $-z_{\textrm{max}} < z < z_{\textrm{max}}$ for some sufficiently large choice of $z_{\textrm{max}}$.
\begin{subequations}
    \label{eq:boundary_value_model_zero_order}%
    \begin{gather}
        \frac{\textrm{d}^2u_0}{\textrm{d}z^2} + c\frac{\textrm{d}u_0}{\textrm{d}z} + u_0\left(1-u_0\right) = 0 \quad \mathrm{ on } \quad -z_{\textrm{max}} < z < 0,
        \label{eq:boundary_value_model_u0}\\
        D\frac{\textrm{d}^2v_0}{\textrm{d}z^2} + c\frac{\textrm{d}v_0}{\textrm{d}z} + \lambda v_0\left(1-v_0\right) = 0 \quad \mathrm{ on } \quad 0 < z <z_{\textrm{max}},
        \label{eq:boundary_value_model_v0}\\
        u_0(-z_{\textrm{max}}) = v_0(z_{\textrm{max}}) = 1,
        \label{eq:boundary_value_z_max_zero_order}\\
        u_0(0) = \, v_0(0) = 0,
        \label{eq:initial_value_0_zero_order}\\
        c = -\kappa_u\;\frac{\textrm{d}u_0(0)}{\textrm{d}z} - \kappa_v\;\frac{\textrm{d}v_0(0)}{\textrm{d}z}.\label{eq:velocity_c_order_zero}
    \end{gather}
\end{subequations}
We treat the boundary value problem for $u_0(z)$ for $z < 0$ and the boundary value problem for $v_0(z)$ for $z > 0$ separately since these problems are only coupled at the moving boundary at $z=0$.  Therefore we consider
\begin{subequations}
    \label{eq:boundary_value_problem_uo}%
    \begin{gather}
        \frac{\textrm{d}^2u_0}{\textrm{d}z^2} + c\frac{\textrm{d}u_0}{\textrm{d}z} + u_0\left(1-u_0\right) = 0 \quad \mathrm{ on } \quad -z_{\textrm{max}} < z < 0,\\
        u_0(-z_{\textrm{max}})=1, \quad u_0(0) =0,
    \end{gather}
\end{subequations}
and
\begin{subequations}
    \label{eq:boundary_value_problem_vo}%
    \begin{gather}
        D\frac{\textrm{d}^2v_0}{\textrm{d}z^2} + c\frac{\textrm{d}v_0}{\textrm{d}z} + \lambda v_0\left(1-v_0\right) = 0 \quad \mathrm{ on } \quad 0 < z < z_{\textrm{max}},\\
        v_0(z_{\textrm{max}}) = 1, \quad v_0(0) = 0.
    \end{gather}
\end{subequations}
separately.  Both Equation (\ref{eq:boundary_value_problem_uo}) and Equation (\ref{eq:boundary_value_problem_vo}) involve an unknown parameter $c$, which we determine later.  The discretization of both boundary value problems is very similar so we present the details for   Equation (\ref{eq:boundary_value_problem_uo}).
Numerical solutions are obtained by discretizing the derivative terms on a uniform discretization of the domain, $z_i = -z_{\textrm{max}} + i\Delta z$, where $\Delta z$ is the constant grid spacing and $i = 0, \dots, N,$ where $N = z_{\textrm{max}}/\Delta z.$  We use $u_{0}^{(i)} = u_0(z_i)$ to denote the discrete approximation of $u_0(z)$ at the $i$th grid point.  Discretizing the boundary value problem at the internal grid points, and the boundary conditions at the boundary points leads to a system of nonlinear algebraic equations given by $\vect{F}(u_{0}^{(i)}) = \vect{0}$ that we solve using Newton-Raphson iteration.  Here,
$\vect{F}$ is given by,
\begin{subequations}
    \label{eq:bvp_u0_discretised}%
    \begin{align}
        F_0 &= u_{0}^{(0)} - 1,\\
        F_i &= \frac{u_0^{(i+1)} - 2u_0^{(i)} + u_0^{(i-1)}}{(\Delta z)^2} + c\frac{u_0^{(i+1)} - u_0^{(i-1)}}{2\Delta z} + u_0^{(i)}\left(1-u_0^{(i)}\right), \quad i = 1, \dots, N-1,\\
        F_{N} &= u_0^{(N)}\;,
    \end{align}
\end{subequations}
We then employ a Newton-Raphson iterative method using some initial estimate $\vect{u}_0^{(0)}$ to give,
\begin{equation}
    \label{eq:newton_method_u0}%
    \vect{u}_0^{(n+1)} = \vect{u}_0^{(n)} - J^{-1}\left(\vect{u}_0^{(n)}\right)\vect{F}\left(\vect{u}_0^{(n)}\right),
\end{equation}
where the superscript $n$ counts the number of iterations and $J_{ij} = \partial_j F_i$ are entries in the Jacobian matrix for $\vect{F}(u_{0}^{(i)})$.  The iterative solver is stopped when we observe $\lVert \vect{u}_0^{(n+1)} - \vect{u}_0^{(n)} \rVert < 1 \times 10^{-6}$.   A similar approach is used to solve the boundary value problem for $v_0(z)$, given by Equation (\ref{eq:boundary_value_problem_vo}).

To estimate $c$ we start with an estimate $c^{(0)}$ and use this value to solve the two boundary value problems for $u_0$ and $v_0$, as described above.  We then discretize the moving boundary condition and define
\begin{equation}
    \label{eq:discretised_Stefan_condition}%
    f^{(k)} := c^{(k)} + \kappa_u\frac{3u_{0}^{(N, k)} - 4u_{0}^{({N-1} ,k)} + u_{0}^{({N-2},k)}}{2\Delta z} + \kappa_v\frac{-3v_{0}^{(0,k)} + 4v_{0}^{(1 ,k)} - v_{0}^{(2 ,k)}}{2\Delta z},
\end{equation}
where $c^{(k)}$ is the current estimate of the wave speed, and $u_{0}^{(i, k)}$ and $v_{0}^{(i, k)}$ for refer to particular values of $u_0$ and $v_0$ on the mesh, adjacent to the moving boundary problem at the $k$th iteration.  To satisfy the moving boundary condition, Equation (\ref{eq:velocity_c_order_zero}) we estimate the value of $c$ that corresponds to $f = 0$ using the univariate Newton-Raphson method,
\begin{equation}
    \label{eq:newton_c}%
    c^{(k+1)} = c^{(k)} - \frac{f\left(c^{(k)}\right)}{f'\left(c^{(k)}\right)}.
\end{equation}
Iterations continue until $\lVert c^{(k+1)} - c^{(k)} \rVert < 1 \times 10^{-6}$, and we estimate $f'$ using a finite difference approximation
\begin{equation}
    \label{eq:f_approx}
    \frac{\textrm{d}f}{\textrm{d}c} = \frac{f(c + \delta c) - f(c)}{\delta c}.
\end{equation}
All results correspond to setting $\delta c = 1 \times 10^{-6}$.

\subsection*{Perturbation solution of the leading order boundary value problem}
Following the approach of El-Hachem et al.~\cite{ElHachem2020} we treat Equation (\ref{eq:leading_order_model}) as two boundary value problems that are coupled through the moving boundary condition at $z=0$.  Therefore, we consider
\begin{subequations}
    \label{eq:bvp_u0_infty}%
    \begin{gather}
        \frac{\textrm{d}^2u_0}{\textrm{d}z^2} + c\frac{\textrm{d}u_0}{\textrm{d}z} + u_0\left(1-u_0\right) = 0 \quad \mathrm{ on } \quad -\infty < z < 0,
        \label{eq:u_0_zero_order}\\
      \lim_{z\to -\infty}  u_0(z)=1, \quad u_0(0) =0,
      \label{eq:boundary_condition_u_0_zero_order}
    \end{gather}
\end{subequations}
and
\begin{subequations}
    \label{eq:bvp_v0_infty}%
    \begin{gather}
        D\frac{\textrm{d}^2v_0}{\textrm{d}z^2} + c\frac{\textrm{d}v_0}{\textrm{d}z} + \lambda v_0\left(1-v_0\right) = 0 \quad \mathrm{ on } \quad 0 < z < \infty,
        \label{eq:v_0_zero_order}\\
        \lim_{z\to -\infty}v_0(z) = 1, \quad v_0(0) = 0,
        \label{eq:boundary_condition_v_0_zero_order}
    \end{gather}
\end{subequations}
separately.  Noting that Equation (\ref{eq:u_0_zero_order}) for $u_0(z)$ is identical to Equation (\ref{eq:v_0_zero_order}) for $v_0(z)$ when $D=\lambda=1$ so it is sufficient for us to consider Equation (\ref{eq:v_0_zero_order}). To proceed we re-write Equation (\ref{eq:v_0_zero_order}) as a first-order system,
\begin{subequations}
    \label{eq:first_order}%
    \begin{align}
    \dfrac{\textrm{d}v_0}{\textrm{d}z}&=x,\\
    \dfrac{\textrm{d}x}{\textrm{d}z} &= -\frac{c}{D} x-\frac{\lambda}{D}\;v_0(1-v_0)\;.
    \end{align}
\end{subequations}
Using the chain rule we rewrite this system as

\begin{equation}
\label{eq:x_v}
\frac{\textrm{d}x}{\textrm{d}v_0}=\frac{-cx-\lambda v_0(1-v_0)}{Dx}.
\end{equation}
While we cannot solve this equation for $x(v_0)$ explicitly, we can make progress by treating $c$ as a small parameter and assuming a series solution of the form $x(v_0)=x_0(v_0)+cx_1(v_0)+\mathcal{O}(c^2)$.  Substituting this expansion into Equation (\ref{eq:x_v}) gives us
\begin{subequations}
    \label{eq:two_terms}%
    \begin{align}
       \frac{\textrm{d}x_0}{\textrm{d}v_0}&=-\frac{\lambda v_0(1-v_0)}{Dx_0},\\
       \frac{\textrm{d}x_1}{\textrm{d}v_0}&=\frac{\lambda x_1 v_0(1-v_0)}{Dx_0^2}-\frac{1}{D},
    \end{align}
\end{subequations}
which can both be solved exactly.  Imposing boundary conditions $x_0(1)=0$ and $x_1(1)=0$ we can write a two-term perturbation solution as
\begin{equation}
\label{eq:perturbation_solution}
x(v_0)=\frac{\textrm{d}v_0}{\textrm{d}z}=\pm\sqrt{\frac{\lambda}{D}\left(-v_0^2+\frac{2v_0^3}{3}+\frac{1}{3}\right)}-c \frac{(2-v_0)\;(1+2v_0)^{3/2}-3\sqrt{3}}{5D(1-v_0)\sqrt{1+2v_0}}+\mathcal{O}(c^2).
\end{equation}
Following El-Hachem et al.~\cite{ElHachem2020}, integrating Equation (\ref{eq:perturbation_solution}) numerically with respect to $z$ using standard numerical ODE tools in the Julia DifferentialEquation.jl package we obtain $v_0(z)$.  If we repeat this process setting $D=\lambda=1$ we obtain the solution for $u_0(z)$.

\begin{figure}
\centering
\includegraphics[width=1.0\textwidth]{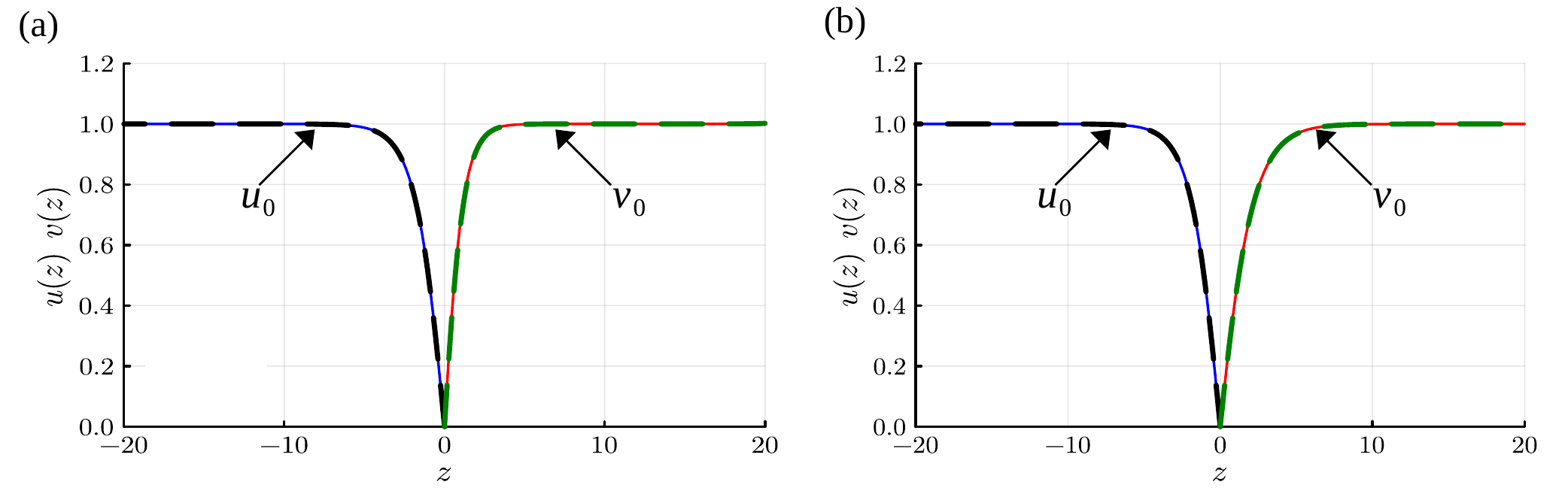}
\caption{Comparison of numerical (solid lines) and perturbation solutions (dashed lines) for $u_0(z)$ and $v_0(z)$.  (a) $\kappa_u=0.4, \kappa_v=0.2$, $D=0.5$, $\lambda=1$ and $c=0.04$. (b)$\kappa_u=0.4, \kappa_v=0.2$, $D=1$, $\lambda=0.5$ and $c=0.11$.}
\label{Figure_10}
\end{figure}

Figure \ref{Figure_10} provides two visual comparisons of $u_0(z)$ and $v_0(z)$ computed using this perturbation approximation and the numerical approach outlined in the previous section.  In both cases we see that the numerical solutions are visually distinguishable from the perturbation solution and so this gives us confidence in our numerical results.  Of course, the perturbation solutions are relevant for $|c| \ll 1$ and if we are interested in travelling wave solutions with larger values of $|c|$ we would use the numerical approach and always take care to ensure that the numerical solution is grid-independent by checking that the solutions are indistinguishable when we refine $\Delta z$.

\newpage
\subsection*{Numerical solution of the $\mathcal{O}(\varepsilon)$ boundary value problems}
The approach for solving Equation (\ref{eq:first_order_model}) for the correction terms follows closely with the numerical method used to solve the leading order problem described above.   The first step is to consider the boundary value problem on a truncated domain,
\begin{subequations}
    \label{eq:boundary_value_model_first_order}
    \begin{gather}
        \frac{\textrm{d}^2{u}_1}{\textrm{d}z^2} + c\frac{\textrm{d}u_1}{\textrm{d}z} + [1-\omega-q^2-2\;u_0]u_1 +(\omega+q^2)\;  \frac{\textrm{d}u_0}{\textrm{d}z}=0 \quad \mathrm{ on } \quad  -z_{\textrm{max}}<z < 0, \label{eq:boundary_value_model_u1}\\
        D\frac{\textrm{d}^2 v_1}{\textrm{d}z^2} + c\frac{\textrm{d} v_1}{\textrm{d}z} + [\lambda-\omega-Dq^2-2\lambda\;v_0] v_1 +(\omega+Dq^2)\;  \frac{\textrm{d}v_0}{\textrm{d}z}=0\quad \mathrm{ on } \quad 0<z<z_{\textrm{max}},  \label{eq:boundary_value_model_v1}\\
       u_1(-z_{\textrm{max}}) = v_1(z_{\textrm{max}}) = u_1(0) =  v_1(0) = 0,  \label{eq:initial_value_z_0_first_order}\\
        \omega = -\kappa_u\;\frac{\textrm{d}u_1(0)}{\textrm{d}z} - \kappa_v\;\frac{\textrm{d}v_1(0)}{\textrm{d}z}.\label{eq:omega_term_first_order}
    \end{gather}
\end{subequations}
As before we treat the boundary value problem for $u_1(z)$ and $v_1(z)$ separately
\begin{subequations}
    \label{eq:bvp_second_order_u1}%
    \begin{gather}
       \frac{\textrm{d}^2 u_1}{\textrm{d}z^2} + c\frac{\textrm{d} u_1}{\textrm{d}z} + [1-\omega-q^2-2\;u] u_1 +(\omega+q^2)\;  \frac{\textrm{d}u_0}{\textrm{d}z}=0 \quad \mathrm{ on } \quad  -z_{\textrm{max}}<z < 0,\\
        u_1(-z_{\textrm{max}})= u_1(0) =0,
    \end{gather}
\end{subequations}
and
\begin{subequations}
    \label{eq:bvp_second_order_v1}%
    \begin{gather}
        D\frac{\textrm{d}^2 v_1}{\textrm{d}z^2} + c\frac{\textrm{d} v_1}{\textrm{d}z} + [\lambda-\omega-Dq^2-2\lambda\;v_0]  v_1 +(\omega+Dq^2)\;  \frac{\textrm{d}v}{\textrm{d}z}=0\quad \mathrm{ on } \quad 0<z<z_{\textrm{max}},\\
        v_1(z_{\textrm{max}})= v_1(0) =0.
    \end{gather}
\end{subequations}
To solve Equation (\ref{eq:bvp_second_order_u1}) we uniformly discretize the truncated domain and approximate all derivative terms using standard finite difference approximations to give
\begin{subequations}
  \label{eq:app_shooting_bvp_fo_fkpp_discretisation}%
  \begin{gather}
    u_1^{(0)} = 0,\label{eq:app_shooting_bvp_fo_fkpp_discretisation_left}\\
    \begin{gathered}
      \frac{u_1^{(i+1)} - 2 u_1^{(i)} + u_1^{(i-1)}}{(\Delta z)^2} + c\frac{u_1^{(i+1)} - u_1^{(i-1)}}{2\Delta z} + \left(1-\omega-q^2-2u_0^{(i)}\right) u_1^{(i)} \\= -\left(\omega + q^2\right)\frac{u_0^{(i+1)} - u_0^{(i-1)}}{2\Delta z},  \quad i = 1, \dots, N-1,
    \end{gathered}\label{eq:app_shooting_bvp_fo_fkpp_discretisation_interior}\\
    u_1^{(N)} = 0, \label{eq:app_shooting_bvp_fo_fkpp_discretisation_right}
  \end{gather}
\end{subequations}
where $c$ and all terms involving $u_0$ and $v_0$ are treated as known outputs from the numerical solution of the leading order problem, and $q$ is the specified wave number. We approximate Equation (\ref{eq:bvp_second_order_v1}) using a very similar approach.  To compute the growth rate we take an initial estimate of $\omega^{(0)}$ and discretize Equation (\ref{eq:omega_term_first_order}),
\begin{equation} \label{eq:omega_condition}%
g^{(k)} = \omega^{(k)} + \kappa_u\dfrac{3 u_{1}^{(N,k)} - 4 u_{1}^{(N-1,k)} + u_{1}^{(N-2,k)}}{2\Delta z} + \kappa_v\dfrac{-3 v_{1}^{(0,k)} + 4 v_{1}^{(1,k)} - v_{1}^{(2,k)}}{2\Delta z},
\end{equation}
and $u_{1}^{(i, k)}$ and $v_{1}^{(i, k)}$ for refer to particular values of $u_1$ and $v_1$ on the mesh, adjacent to the moving boundary problem at the $k$th iteration.  Implementing the one-variable Newton-Raphson method we solve $g(\omega)=0$ to give an improved estimate of $\omega$,
\begin{equation}
    \label{eq:newton_omega}%
    \omega^{(k+1)} = \omega^{(k)} - \frac{g\left(\omega^{(k)}\right)}{g'\left(\omega^{(k)}\right)},
\end{equation}
where we estimate $g'$ using a finite difference approximation
\begin{equation}
    \label{eq:f_approx_g}
    \frac{\textrm{d}g}{\textrm{d}\omega} = \frac{g(\omega + \delta \omega) - g(\omega)}{\delta \omega}.
\end{equation}
All results correspond to setting $\delta \omega = 1 \times 10^{-6}$.


\newpage
\begin{thebibliography}{99}
\bibitem{Murray2002} Murray JD (2002) Mathematical biology I: an introduction. Springer.

\bibitem{Kot2003} Kot M (2003) Elements of Mathematical Ecology. Cambridge University Press, Cambridge.

\bibitem{Edelstein2005} Edelstein-Keshet L (2005) Mathematical Models in Biology. SIAM, Philadelphia.

\cbl
\bibitem{Keitt2001} Keitt TH, Lewis MA, Holt RD. 2002.  Allee effects, invasion pinning, and species’ borders.  The American Naturalist. 157, 203--216. (\href{https://doi.org/10.1086/318633}{10.1086/318633}). \cb

\cbl
\bibitem{Taylor2005} Taylor CM, Hastings A (2005) Allee effects in biological invasions.  Ecology Letters. 8, 895-908. (\href{ https://doi.org/10.1111/j.1461-0248.2005.00787.x}{doi:10.1111/j.1461-0248.2005.00787.x}). \cb


\cbl
\bibitem{Barton2011} Barton NH,  Turelli M. 2011. Spatial waves of advance with bistable dynamics: cytoplasmic and genetic analogues of Allee effects. The American Naturalist. 178, E48–E75. (\href{https://doi.org/10.1086/661246}{10.1086/661246}). \cb

\cbl
\bibitem{Fadai2020} Fadai NT, Simpson MJ (2020) Population dynamics with threshold effects give rise to a diverse family of Allee effects. Bulletin of Mathematical Biology.  82,  74. (\href{https://doi.org/10.1007/s11538-020-00756-5}{10.1007/s11538-020-00756-5}). \cb

\cbl
\bibitem{Surendran2020} Surendran A, Plank MJ, Simpson MJ. 2020 Population dynamics with spatial structure and an Allee effect. Proceedings of the Royal Society A.  476, 20200501. (\href{http://doi.org/10.1098/rspa.2020.0501}{10.1098/rspa.2020.0501}). \cb


\bibitem{Lewis1993} Lewis MA, Kareiva P. 1993. Allee dynamics and the spread of invading organisms. Theoretical Population Biology. 43, 141–158. (\href{http://dx.doi.org/10.1006/tpbi.1993.1007}{doi:10.1006/tpbi.1993.1007}).

\bibitem{Li2022} Li Y, Johnston ST, Buenzli PR, van Heijster P, Simpson MJ. 2022.  Extinction of bistable populations is affected by the shape of their initial spatial distribution.   Bulletin of Mathematical Biology. 84, 21. (\href{https://doi.org/10.1007/s11538-021-00974-5}{doi:10.1007/s11538-021-00974-5}).

\cbl
\bibitem{Cantrell1998} Cantrell RS, Cosner C. 1998. On the effects of spatial heterogeneity on the persistence of interacting species.  Journal of Mathematical Biology. 37, 103–-145.  (\href{https://doi.org/10.1007/s002850050122}{10.1007/s002850050122}). \cb


\cbl \bibitem{Maciel2013} Maciel GA, Lutscher F. 2013. How individual movement response to habitat edges affects population persistence and spatial spread. The American Naturalist. 182, 42--52.  (\href{https://doi.org/10.1086/670661}{10.1086/670661}).
\cb

\cbl
\bibitem{Fussell2019} Fussell EF, Krause AL, Van Gorder RA. 2019. Hybrid approach to modeling spatial dynamics of systems with generalist predators. Journal of Theoretical Biology. 462, 26--47. (\href{https://doi.org/10.1016/j.jtbi.2018.10.054}{10.1016/j.jtbi.2018.10.054}).\cb


\cbl
\bibitem{Gatenby1996} Gatenby RA, Gawlinski ET. 1996. A reaction-diffusion model of cancer invasion. Cancer Research. 56, 5745--5753. (\href{https://aacrjournals.org/cancerres/article/56/24/5745/502885/A-Reaction-Diffusion-Model-of-Cancer-Invasion}{Cancer Research 56}). \cb


\cbl
\bibitem{Schofield2011} Schofield JW, Gaffney EA, Gatenby RA, Maini PK. 2011.  Tumour angiogenesis: The gap between theory and experiments. Journal of Theoretical Biology. 274, 97--102. (\href{https://doi.org/10.1016/j.jtbi.2011.01.012}{10.1016/j.jtbi.2011.01.012}). \cb

\cbl
\bibitem{Browning2019} Browning AP, Haridas P, Simpson MJ. 2019.  A Bayesian sequential learning framework to parameterise continuum models of melanoma invasion into human skin. Bulletin of Mathematical Biology. 81, 676--698. (\href{https://doi.org/10.1007/s11538-018-0532-1}{10.1007/s11538-018-0532-1}). \cb

\bibitem{Fisher1937} Fisher RA (1937) The wave of advance of advantageous genes. Annals of Eugenics.  7:355--369.

\bibitem{Kolmogorov1937} Kolmogorov A, Petrovsky I, Piskunov N. 1937. A study of the equation of diffusion with increase in the quantity of matter, and its application to a biological problem. Moscow University Mathematics Bulletin. 1, 1-26.

\bibitem{Skellam1951} Skellam JG. 1951. Random dispersal in theoretical populations. Biometrika. 38, 196-218. (\href{https://doi.org/10.1093/biomet/38.1-2.196}{doi:10.1093/biomet/38.1-2.196}).

\bibitem{Maini2004} Maini PK, McElwain  DLS, Leavesley DI (2004) Traveling wave model to interpret a wound--healing cell migration assay for human peritoneal mesothelial cells. Tissue Engineering. 10:475--482. (\href{https://doi.org/10.1089/107632704323061834}{doi:10.1089/107632704323061834}).

\bibitem{Sengers2007} Sengers BG, Please CP, Oreffo ROC. 2007. Experimental characterization and computational modelling of two-dimensional cell spreading for skeletal regeneration. Journal of The Royal Society Interface. 4, 1107-1117. (\href{https://doi.org/10.1098/rsif.2007.0233}{doi:10.1098/rsif.2007.0233}).

\bibitem{Jin2016} Jin W, Shah ET, Penington CJ, McCue SW, Chopin LK, Simpson MJ (2016) Reproducibility of scratch assays is affected by the initial degree of confluence: experiments, modelling and model selection. Journal of Theoretical Biology. 390: 136--145.  (\href{https://doi.org/10.1016/j.jtbi.2015.10.040}{doi:10.1016/j.jtbi.2015.10.040}).

\bibitem{Jin2018} Jin W, Lo K-Y, Chou S-E, McCue SW, Simpson MJ (2018) The role of initial geometry in experimental models of wound closing. Chemical Engineering Science. 179:221--226. (\href{https://doi.org/10.1016/j.ces.2018.01.004}{doi:10.1016/j.ces.2018.01.004}).

\bibitem{McCue2019} McCue SW, Jin W,  Moroney TJ, Lo K--Y, Chou SE, Simpson MJ (2019) Hole--closing model reveals exponents for nonlinear degenerate diffusivity functions in cell biology. Physica D: Nonlinear Phenomena. 398:130--140. (\href{https://doi.org/10.1016/j.physd.2019.06.005}{doi:10.1016/j.physd.2019.06.005}).

\bibitem{ElHachem2019} El-Hachem, M, McCue SW, Jin W, Du Y, Simpson MJ. 2019. Revisiting the Fisher--Kolmogorov--Petrovsky--Piskunov equation to interpret the spreading\textendash extinction dichotomy. Proceedings of the Royal Society A. 475, 20190378. (\href{https://doi.org/10.1098/rspa.2019.0378}{doi:10.1098/rspa.2019.0378}).

\bibitem{Crank1987} Crank J. 1987. Free and Moving Boundary Problems. Oxford University Press.

\bibitem{Gupta2017} Gupta SC. 2017. The classical Stefan problem: basic concepts, modelling and analysis with quasi-analytical solutions and methods. Elsevier.

\bibitem{Du2010} Du Y, Lin Z. 2010. Spreading-vanishing dichotomy in the diffusive logistic model with a free boundary. SIAM Journal on Mathematical Analysis. 44, 377-405. (\href{https://doi.org/10.1137/090771089}{doi:10.1137/090771089}).

\bibitem{BrosaPlanella2019} Brosa Planella F, Please CP, Van Gorder RA. 2019. Extended Stefan problem for solidification of binary alloys in a finite planar domain. SIAM Journal on Applied Mathematics. 79, 876-913. (\href{https://doi.org/10.1137/18M118699X}{doi:10.1137/18M118699X}).

\bibitem{BrosaPlanella2021} Brosa Planella F, Please CP, Gorder RA. 2021. Extended Stefan problem for the solidification of binary alloys in a sphere. European Journal of Applied Mathematic. 32, 242-279.  (\href{https://doi.org/10.1017/S095679252000011X}{doi:10.1017/S095679252000011X}).

\bibitem{Mitchell2009} Mitchell SL, Vynnycky M. 2009.  Finite-difference methods with increased accuracy and correct initialization for one-dimensional Stefan problems. Applied Mathematics and Computation. 215, 1609-1621. (\href{http://dx.doi.org/10.1016/j.amc.2009.07.054}{doi:10.1016/j.amc.2009.07.054}).

\bibitem{Mitchell2010} Mitchell SL, Myers TG. 2010. Improving the accuracy of heat balance integral methods applied to thermal problems with time dependent boundary conditions. International Journal of Heat and Mass Transfer. 53, 3540-3551. (\href{http://dx.doi.org/10.1016/j.ijheatmasstransfer.2010.04.015}{doi:10.1016/j.ijheatmasstransfer.2010.04.015})

\bibitem{Dalwadi2020} Dalwadi MP, Waters SL, Byrne HM, Hewitt IJ. 2020.  A mathematical framework for developing freezing protocols in the cryopreservation of cells. SIAM Journal on Applied Mathematics. 80, 657-689. (\href{http://dx.doi.org/10.1137/19M1275875}{doi:10.1137/19M1275875})

\bibitem{Ward1997}  Ward JP, King JR. 1997. Mathematical modelling of avascular-tumour growth. Mathematical Medicine and Biologyl. 14, 39-69. (\href{https://doi.org/10.1093/imammb/14.1.39}{doi:10.1093/imammb/14.1.39}).

\bibitem{Gaffney1999} Gaffney EA, Maini PK, McCaig CD, Zhao M, Forrester JV. 1991. Modelling corneal epithelial wound closure in the presence of physiological electric fields via a moving boundary formalism. Mathematical Medicine and Biology: A Journal of the IMA. 16, 369–393. (\href{http://dx.doi.org/10.1093/imammb/16.4.369}{doi:10.1093/imammb/16.4.369}).

\bibitem{Shuttleworth2019} Shuttleworth R, Trucu D. 2019. Multiscale modelling of fibres dynamics and cell adhesion within moving boundary cancer invasion. Bulletin of Mathematical Biology. 81, 2176-2219. (\href{https://doi.org/10.1007/s11538-019-00598-w}{doi:10.1007/s11538-019-00598-w}).

\bibitem{ElHachem2021} El-Hachem M, McCue SW,  Simpson MJ. 2021. Invading and receding sharp-fronted travelling waves. Bulletin of Mathematical Biology. 83, 35. (\href{https://doi.org/10.1007/s11538-021-00862-y}{doi:10.1007/s11538-021-00862-y}).

\bibitem{Simpson2020} Simpson MJ. 2020. Critical length for the spreading–vanishing dichotomy in higher dimensions. ANZIAM Journal. 62(1), 3-17. (\href{https://doi.org/10.1017/S1446181120000103}{doi:10.1017/S1446181120000103})

\bibitem{ElHachem2020} El-Hachem M, McCue SW,  Simpson MJ. 2020. A sharp-front moving boundary model for malignant invasion. Physica D: Nonlinear Phenomena. 412, 132639. (\href{https://doi.org/10.1016/j.physd.2020.132639}{doi:10.1016/j.physd.2020.132639}).

\bibitem{Haridas2017} Haridas P, McGovern JA, McElwain DLS, Simpson MJ. 2017. Quantitative comparison of the spreading and invasion of radial growth phase and metastatic melanoma cells in a three-dimensional human skin equivalent model. PeerJ. 5, e3754-e3754. (\href{https://doi.org/10.7717/peerj.3754}{doi:10.7717/peerj.3754}).

\bibitem{Tam2022}  Tam AKY, Simpson MJ. 2022. The effect of geometry on survival and extinction in a moving-boundary problem motivated by the Fisher--KPP equation. Physica D: Nonlinear Phenomena. 438, 133305. (\href{https://doi.org/10.1016/j.physd.2022.133305}{doi:10.1016/j.physd.2022.133305})

\bibitem{Tam2023} Tam AKY, Simpson MJ. 2023. Pattern formation and front stability for a moving-boundary model of biological invasion and recession. Physica D: Nonlinear Phenomena. 444, 133593. (\href{https://doi.org/10.1016/j.physd.2022.133593}{doi:10.1016/j.physd.2022.133593}).

\bibitem{Sethian1999} Sethian JA. 1999. Level set methods and fast marching methods: evolving interfaces in computational geometry, fluid mechanics, computer vision, and materials science. Cambridge University Press.

\bibitem{Osher2003} Osher S, Fedkiw RP. 2003. Level set methods and dynamic implicit surfaces. Springer-Verlag.

\bibitem{Aslam2004} Aslam TD. 2004. A partial differential equation approach to multidimensional extrapolation. Journal of Computational Physics. 193, 349-355. (\href{https://doi.org/10.1016/j.jcp.2003.08.001}{doi:10.1016/j.jcp.2003.08.001}).

\bibitem{Simpson2005} Simpson MJ, Landmann KA, Clemen, TP. 2005. Assessment of a non-traditional operator split algorithm for simulation of reactive transport. Mathematics and Computers in Simulation. 70, 44-60. (\href{https://doi.org/10.1016/j.matcom.2005.03.019}{doi:10.1016/j.matcom.2005.03.019}).

\bibitem{Rackauckas2017} Rackauckas CV, Nie Q. 2017. DifferentialEquations.jl A performant and feature-rich ecosystem for solving differential equations in Julia.  Journal of Open Research Software. 5, 15. (\href{https://doi.org/10.5334/jors.151}{10.5334/jors.151}).

\bibitem{Yang2002} Yang J, D'Onofrio A, Kalliadasis S, De Wit A. 2002. Rayleigh--Taylor instability of reaction--diffusion acidity fronts. Journal of Chemical Physics.  117, 9395-9408. (\href{https://doi.org/10.1063/1.1516595}{doi:10.1063/1.1516595}).

\bibitem{Straint1988} Straint J. 1988. Linear stability of planar solidification fronts. Physica D: Nonlinear Phenomena. 30, 297-320. (\href{https://doi.org/10.1016/0167-2789(88)90023-1}{doi:10.1016/0167-2789(88)90023-1}).

\bibitem{Muller2002} Muller J, van Saarloos W. 2002. Morphological instability and dynamics of fronts in bacterial growth models with nonlinear diffusion.  Physical Review E. 65, 061111. (\href{https://doi.org/10.1103/PhysRevE.65.061111}{10.1103/PhysRevE.65.061111}).

\bibitem{Chen1997} Chen S, Merriman B, Osher S, Smereka P. 1997. A simple level set method for solving Stefan problems. Journal of Computational Physics. 135, 8-29.  (\href{https://doi.org/10.1006/jcph.1997.5721}{10.1006/jcph.1997.5721}).

\bibitem{Harabetian1998} Harabetian E, Osher S. 1998. Regularization of ill-posed problems via the level set approach.  SIAM Journal on Applied Mathematics. 58, 1689-1706.  (\href{https://doi.org/10.1137/S003613999529079}{10.1137/S003613999529079}).

\bibitem{Chadam1983} Chadam J, Ortoleva P. 1983. The stabilizing effect of surface tension on the development of the free boundary in a planar, one-dimensional, Cauchy-Stefan problem. IMA Journal of Applied Mathematics. 30, 57-66. (\href{https://doi.org/10.1093/imamat/30.1.57}{doi:10.1093/imamat/30.1.57})


\bibitem{Sethian1992} Sethian JA, Straint J. 1992. Crystal growth and dendritic solidification.  Journal of Computational Physics.  98, 231-253.   (\href{https://doi.org/10.1016/0021-9991(92)90140-T}{10.1016/0021-9991(92)90140-T}).

\bibitem{Morrow2019} Morrow L, Moroney T, McCue SW. 2019. Numerical investigation of controlling interfacial instabilities in non-standard Hele-Shaw configurations.  Journal of Fluid Mechanics. 877, 1063--1097. (\href{https://doi.org/10.1017/jfm.2019.623}{10.1017/jfm.2019.623}).

\bibitem{King2005} King JR, Evans JD. 2005. Regularization by kinetic undercooling of blowup in the ill-posed Stefan problem. SIAM Journal on Applied Mathematics. 65, 1677-1707.  (\href{https://doi.org/10.1137/04060528X}{doi:10.1137/04060528X}).

\bibitem{Sherratt1990}Sherratt JA, Murray JD. 1990. Models of epidermal wound healing. Proceedings of the Royal Society B: Biological Sciences. 241, 29-36. (\href{https://doi.org/10.1098/rspb.1990.0061}{doi:10.1098/rspb.1990.0061}).

\bibitem{Tam2018}Tam AKY, Green JEF, Balasuriya S, Tek EL, Gardner JM, Sundstrom JF, Jiranek V, Binder BJ. 2018. Nutrient-limited growth with non-linear cell diffusion as a mechanism for floral pattern formation in yeast biofilms. Journal of Theoretical Biology. 448, 122-141. (\href{https://doi.org/10.1016/j.jtbi.2018.04.004}{10.1016/j.jtbi.2018.04.004}).

\end{thebibliography}
\end{document}